\documentclass{jpp}
\usepackage{graphicx}

\usepackage[utf8]{inputenc}
\usepackage[T1]{fontenc}
\usepackage{amsmath}
\usepackage{subcaption}
\usepackage{hyperref}
\hypersetup{
     colorlinks   = true,
     citecolor    = blue,
     linkcolor    = blue}

\renewcommand{\v}[1]{\ensuremath{\mbox{\boldmath$ #1 $}}} 
 
\newcommand{\uv}[1]{\ensuremath{\mathbf{\hat{#1}}}} 
\newcommand{\avg}[1]{\left< #1 \right>} 
\renewcommand{\d}[2]{\frac{d #1}{d #2}} 
\newcommand{\pd}[2]{\frac{\partial #1}{\partial #2}} 
\newcommand{\pdd}[2]{\frac{\partial^2 #1}{\partial #2^2}} 
 
\newcommand{\grad}[1]{\nabla #1} 
\newcommand{\gradv}[1]{\nabla_{{\scalebox{.7}{$\scriptscriptstyle\v{v}$}}} #1} 
\newcommand{\divv}[1]{\nabla_{{\scalebox{.7}{$\scriptscriptstyle\v{v}$}}} \cdot #1} 
\newcommand{\lapv}[1]{\nabla_{{\scalebox{.7}{$\scriptscriptstyle\v{v}$}}}^2 #1} 
\newcommand{\kron}[2]{\delta_{#1 #2}} 

\newcommand{\sind}{s}
\newcommand{\rind}{r}
\newcommand{\sr}{\sind\rind}

\newcommand{\fs}{f_\sind}
\newcommand{\fr}{f_\rind}
\newcommand{\fsr}{f_{\sind,\rind}}
\newcommand{\fsZ}{f_{\sind0}}

\newcommand{\us}{u_{\sind}}

\newcommand{\usi}{u_{\sind,i}}

\newcommand{\ur}{u_{\rind}}
\newcommand{\uri}{u_{\rind,i}}

\newcommand{\vts}{v_{t,\sind}}
\newcommand{\vtr}{v_{t,\rind}}
\newcommand{\vus}{\v{u}_\sind}
\newcommand{\vur}{\v{u}_\rind}
\newcommand{\ms}{m_\sind}
\newcommand{\mr}{m_\rind}
\newcommand{\me}{m_e}
\newcommand{\mi}{m_i}

\newcommand{\qs}{q_\sind}
\newcommand{\qr}{q_\rind}
\newcommand{\upars}{u_{\parallel\sind}}
\newcommand{\uparr}{u_{\parallel\rind}}
\newcommand{\upare}{u_{\parallel e}}
\newcommand{\upari}{u_{\parallel i}}

\renewcommand{\ns}{n_\sind}
\newcommand{\nr}{n_\rind}
\newcommand{\Ts}{T_\sind}
\newcommand{\Tr}{T_\rind}

\newcommand{\MZs}{M_{0,\sind}}
\newcommand{\MOis}{M_{1i,\sind}}
\newcommand{\MOs}{M_{1,\sind}}
\newcommand{\MTs}{M_{2,\sind}}
\newcommand{\MZr}{M_{0,\rind}}
\newcommand{\MOir}{M_{1i,\rind}}
\newcommand{\MOr}{M_{1,\rind}}
\newcommand{\MTr}{M_{2,\rind}}
\newcommand{\MOpars}{M_{1\parallel,\sind}}
\newcommand{\MOparr}{M_{1\parallel,\rind}}
\newcommand{\vu}{\v{u}}
\newcommand{\vue}{\v{u}_e}
\newcommand{\vui}{\v{u}_i}
\newcommand{\vte}{v_{t,e}}
\newcommand{\vti}{v_{t,i}}

\newcommand{\MZstars}{\MZs^\star}
\newcommand{\MOstars}{\MOs^\star}
\newcommand{\MTstars}{\MTs^\star}
\newcommand{\MZstarr}{\MZr^\star}
\newcommand{\MOstarr}{\MOr^\star}
\newcommand{\MTstarr}{\MTr^\star}
\newcommand{\MOparstars}{\MOpars^\star}
\newcommand{\MOparstarr}{\MOparr^\star}

\newcommand{\nusr}{\nu_{\sind\rind}}
\newcommand{\nurs}{\nu_{\rind\sind}}
\newcommand{\usr}{u_{\sind\rind}}
\newcommand{\usri}{u_{\sind\rind,i}}

\newcommand{\urs}{u_{\rind\sind}}
\newcommand{\ursi}{u_{\rind\sind,i}}
\newcommand{\vusr}{\v{u}_{\sind\rind}}

\newcommand{\vtsr}{v_{t,\sind\rind}}
\newcommand{\vtrs}{v_{t,\rind\sind}}
\newcommand{\vuei}{\v{u}_{ei}}
\newcommand{\vuie}{\v{u}_{ie}}
\newcommand{\vtei}{v_{t,ei}}
\newcommand{\vtie}{v_{t,ie}}
\newcommand{\nuei}{\nu_{ei}}
\newcommand{\nuie}{\nu_{ie}}

\newcommand{\uparsr}{u_{\parallel\sind\rind}}
\newcommand{\uparrs}{u_{\parallel\rind\sind}}

\newcommand{\vdim}{{d_v}}
\newcommand{\alphae}{\alpha_{E}}
\newcommand{\fnue}{\delta_e}
\newcommand{\fnui}{\delta_i}
\newcommand{\fnus}{\delta_\sind}

\newcommand{\mrDms}{\frac{\mr}{\ms}}
\newcommand{\msDmr}{\frac{\ms}{\mr}}
\newcommand{\meDmi}{\frac{\me}{\mi}}
\newcommand{\miDme}{\frac{\mi}{\me}}
\newcommand{\TrDTs}{\frac{\Tr}{\Ts}}

\newcommand{\cssr}{c_{s,\sind\rind}}
\newcommand{\boltzH}{\ensuremath{H}}
\newcommand{\entr}{\mathcal{S}}
\newcommand{\vx}{\v{x}}
\newcommand{\vv}{\v{v}}

\newcommand{\vi}{v_i}

\newcommand{\vJsr}{\v{J_{\sind\rind}}}

\newcommand{\logLambdasr}{\log\Lambda_{\sind\rind}}
\newcommand{\intvdim}{\int_{\vdim}}
\newcommand{\dotTsr}{\dot{\mathcal{T}}_{\sind\rind}}
\newcommand{\dotNsr}{\dot{\mathcal{N}}_{\sind\rind}}
\newcommand{\vb}{\v{b}}
\newcommand{\bi}{b_i}
\newcommand{\vh}{\v{h}}

\newcommand{\vhO}{\v{h_0}}
\newcommand{\mcJ}{\mathcal{J}}
\newcommand{\dx}{\mathrm{d}x}
\newcommand{\dv}{\mathrm{d}v}
\newcommand{\dxdv}{\dx\,\dv}
\newcommand{\vpar}{v_\parallel}
\newcommand{\dvpar}{\mathrm{d}\vpar}
\newcommand{\dmu}{\mathrm{d}\mu}

\newcommand{\Nx}{N_x}
\newcommand{\Nv}{N_v}
\newcommand{\numB}{N_b}
\newcommand{\numBx}{N_b^x}
\newcommand{\Lx}{L_x}
\newcommand{\Lv}{L_v}
\newcommand{\Gs}{G_\sind}
\newcommand{\Gr}{G_\rind}
\newcommand{\jmax}{j_{\max}}
\newcommand{\jmin}{j_{\min}}
\newcommand{\kmax}{k_{\max}}
\newcommand{\kmin}{k_{\min}}
\newcommand{\vmax}{v_{\max}}
\newcommand{\vmin}{v_{\min}}
\newcommand{\vimax}{v_{i,\max}}
\newcommand{\vimin}{v_{i,\min}}

\newcommand{\vparmax}{v_{\parallel\max}}
\newcommand{\vparmin}{v_{\parallel\min}}
\newcommand{\mumax}{\mu_{\max}}
\newcommand{\mumin}{\mu_{\min}}
\newcommand{\vSqBar}{\overline{v^2}}
\newcommand{\vparSqBar}{\overline{\vpar^2}}
\newcommand{\vcen}{\check{v}}

\newcommand{\vparcenj}{\check{v}_{\parallel j}}

\newcommand{\gene}{\ensuremath{\mathtt{GENE}}}
\newcommand{\gkeyll}{\ensuremath{\mathtt{Gkeyll}}}

\newenvironment{eqnal}{\equation\aligned}{\endaligned\endequation}

\newcommand{\ignore}[1]{}  

\shorttitle{Improved multispecies Dougherty collisions}
\shortauthor{M. Francisquez, J. Juno, A. Hakim, G. W. Hammett, D. R. Ernst}

\title{Improved multispecies Dougherty collisions}

\author{Manaure Francisquez\aff{1,3}
  \corresp{\email{mfrancis@pppl.gov}},
  James Juno\aff{2}, Ammar Hakim\aff{1}, Greg~W.~Hammett\aff{1} \and Darin~R.~Ernst\aff{3}}

\affiliation{\aff{1}Princeton Plasma Physics Laboratory, Princeton, NJ 08543, USA.
\aff{2}Department of Physics \& Astronomy, University of Iowa, Iowa City, IA 52242, USA.
\aff{3}MIT Plasma Science and Fusion Center, Cambridge, MA 02139, USA.}

\begin{document}

\maketitle

\begin{abstract}
The Dougherty model Fokker-Planck operator is extended to describe nonlinear full-$f$ collisions between multiple species in plasmas. Simple relations for cross-species primitive moments are developed which obey conservation laws, and reproduce familiar velocity and temperature relaxation rates. This treatment of multispecies Dougherty collisions, valid for arbitrary mass ratios, avoids unphysical temperatures and satisfies the \boltzH-theorem unlike an analogous Bhatnagar-Gross-Krook operator. Formulas for both a Cartesian velocity-space and a gyroaveraged operator are provided for use in Vlasov as well as long-wavelength gyrokinetic models. We present an algorithm for the discontinuous Galerkin discretization of this operator, and provide results from relaxation and Landau damping benchmarks.
\end{abstract}

\section{Introduction} \label{sec:intro}

Collisions play an important role in many laboratory and astrophysical plasma processes of interest. They offer a velocity-space dissipative channel in kinetic turbulence and modify transport in fusion devices, to mention a couple. In continuum kinetic models for plasmas, where small-angle collisions prevail, the effect of collisions is incorporated by the Fokker-Planck operator (FPO)~\citep{Rosenbluth1957}. The gyrokinetic form of this operator also exists~\citep{Li2011,Hirvijoki2017,Jorge2019a,Pan2019} and has been shown to agree closely with `model' operators in some parameter ranges~\citep{Pan2020}, but it can also produce significantly different results in others, particularly for instabilities and turbulence driven by the electron temperature gradient~\citep{Pan2021}. Nevertheless, exact FPOs often prove to be analytically and numerically challenging for certain applications. Thus, there is still great interest in using simpler `model' collision operators, several of which have arisen in the last several years~\citep{Abel2008,Sugama2009,Esteve2015,Sugama2019,Frei2021}. These model operators compromise accurate physics for tractability of calculations. Yet these approaches may still have sufficient complexity to deter their use, and mostly exist in linearized form for use in $\delta f$ studies (e.g.~\cite{Kolesnikov2010}).

The FPO's drag and diffusion terms appear in terms of per unit time increments $\langle\Delta v_i\rangle_\sind$ and $\langle \Delta v_i\Delta v_j\rangle_\sind$. A particularly convenient choice is $\langle \Delta v_i \rangle_\sind = -\sum_\rind\nusr (v_i-\usri)$ and $\langle \Delta v_i \Delta v_j \rangle_\sind = 2\sum_\rind\nusr \vtsr^2\kron{i}{j}$, $\nusr$ being a suitably chosen collision frequency ($i=\{1,\dots,\vdim\}$ labels the velocity component in $\vdim$-dimensional velocity-space). This approximation leads to the simple model Fokker-Planck operator
\begin{equation} \label{eq:LBDeq}
\left(\d{\fs}{t}\right)_c = \sum_\rind\nusr \divv{\left[\left(\vv-\vusr\right)\fs+\vtsr^2\gradv{\fs}\right]}.
\end{equation}
For self-species collisions $\vusr=\vus$ and $\vtsr^2=\vts^2=\Ts/\ms$ are the flow velocity and the squared thermal speed of species $\sind$, defined in terms of the velocity moments of the distribution function ($\MZs,~\MOis,~\MTs$) as
\begin{eqnal}
\usi\MZs &= \MOis, \\
\usi\MOis + \vdim\vts^2\MZs &= \MTs,
\end{eqnal}
with such moments given by
\begin{eqnal} \label{eq:moms}
\MZs &= \int^\infty_{-\infty}\fs\,\text{d}^\vdim v, \\
\MOis &= \int^\infty_{-\infty}v_i\fs\,\text{d}^\vdim v, \\
\MTs &= \int^\infty_{-\infty}v^2\fs\,\text{d}^\vdim v.
\end{eqnal}
This frequently used model goes by the various appellations of Kirkwood, Lenard-Bernstein or Dougherty operator. We refer to it as the LBO for simplicity. Its nonlinearity is implicit, since the primitive moments $\usri$ and $\vtsr^2$ are themselves functions of the moments of $\fsr$. We also restrict ourselves to the case of velocity independent collisionality; improvements that retain this additional complexity will be explored in the future. The result is then a tractable operator which, owing to its simplicity, conservative properties, and similarity to the full FPO, is used in numerous kinetic plasma models and, with appropriate modifications, virtually every gyrokinetic model.

These attributes also make it an attractive choice for multispecies collisions. Analytic and computational studies have used Dougherty electron-ion collisions for several decades to the present day~\citep{Ong1970,Pan2018,Shi2019}. This trend however has not established the most appropriate choice of cross velocities and thermal speed, $\usri$ and $\vtsr$. A study of the universal instability, for example, used $\usri=\left[\mr\nr/(\ms\ns)\right]\uri$ and $\vtsr=\vts$, where $\ns$ is the number density of species $\sind$~\citep{Ong1970}, while a separate analysis of ion-acoustic and drift waves later employed $\usri=\left[\nurs\mr/(\nusr\ms)\right]\usi$~\citep{Ong1973}. \cite{Dougherty1967} had proposed a linearized multispecies version of the eponymous operator with $\usri=\uri$ and $\vtsr^2=\left[\mr/\left(\ms+\mr\right)\right]\left(\vts^2+\vtr^2\right)$. More recently~\cite{Jorge2018} chose $\usri=\usi$ and $\vtsr=\vts$ for exploring drift-waves at arbitrary collisionality. Adding to the variance of choices, $\vuei=\vui$ and $\vtei^2=\vte^2+\left(\vui-\vue\right)^2/3$ were assumed in \gene~and \gkeyll~full-$f$ gyrokinetic simulations of LAPD and NSTX~\citep{Pan2018,Shi2019}. Furthermore, the choice of $\usr$ and $\vtsr$ is related to the adoption of a particular collision frequency $\nusr$, which~\cite{Dougherty1964} and other works left unspecified, though~\cite{Dougherty1967} show one possible choice for the linearized operator.

There has thus been a prolonged, non-systematic spread in the choice of cross-species primitive moments, $\usri$ and $\vtsr$, for multispecies collisions with the Dougherty operator. While some of the choices listed above are intuitive and appropriate in some limits, the goal of this manuscript is to more rigorously determine such cross-species primitive moments.~\cite{Greene1973} for example imposed momentum and energy conservation in electron-ion collisions with a Bhatnagar-Gross-Krook (BGK) operator~\citep{Bhatnagar1954}, and required that the cross-species velocity and temperature relaxation rates match those given by the Boltzmann collision integral for Maxwellian distributions: the Morse relaxation rates~\citep{Morse1963}. This procedure yields relations for the $\usri$ and $\vtsr$ needed by the multispecies BGK model. Unfortunately, for unequal massses the resulting formulae can prescribe a negative $\vtsr^2$ as the relative drift $|\usi-\uri|$ increases. It has also been pointed out that, though conservative, this multispecies BGK operator cannot be proven to have (or not have) an \boltzH-theorem~\citep{Haack2017}.

In what follows we present three different approaches to determining the Dougherty cross-species primitive moments $\usr$ and $\vtsr$, drawing from the ideas of~\cite{Greene1973} and~\cite{Haack2017} employed for the BGK operator. We begin with the presentation of these approaches in the context of Vlasov-Maxwell models (section~\ref{sec:theory}). The proposed multi-species  full-$f$ nonlinear Dougherty operator is also shown to not decrease the entropy. Entropy production stands as a challenging constraint in some other collision models. For example, a modern linear $\delta f$ formulation of multi-species collisions only has an \boltzH-theorem when temperatures are equal~\citep{Sugama2019}. Furthermore, there is little work on full-$f$ collision models; one such operator presented by~\cite{Esteve2015} has been linearized and is also only able to satisfy the \boltzH-theorem for equal temperatures. Section~\ref{sec:theory} ends with a provision of equivalent formulas for the gyroaveraged Dougherty operator which is frequently used in long-wavelength gyrokinetic simulations~\citep{Francisquez2020}. These formulas are then implemented in the discontinuous Galerkin code~\cite{gkeyllWeb} using an algorithm described in section~\ref{sec:algorithm}. Then, section~\ref{sec:results} provides a series of Vlasov and gyrokinetic benchmarks illustrating the conservative properties of the algorithm and the differences between the three different strategies for selecting multispecies primitive moments. We also provide a benchmark comparing the Landau damping rate of electron Lagnmuir waves with the multispecies Dougherty operator against the results using a Fokker-Planck operator. Concluding remarks are provided in section~\ref{sec:conclusion}.

\section{Multispecies Dougherty operators} \label{sec:theory}

In this section we provide three different sets of formulas for the cross-species primitive moments in the LBO. The first is analogous to Greene's treatment of the BGK operator~\citep{Greene1973} and we therefore name it the LBO-G. It introduces a free parameter that is insufficiently constrained at present. We thus complement that approach with the ideas of~\cite{Haack2017}, where two different BGK operators were proposed which independently attempt to match the FPO's momentum and thermal relaxation rates. These are the LBO-EM and LBO-ET, respectively (these operators were also recently implemented in the GENE-X code~\citep{Ulbl2021}). We conclude this section with similar formulas for a gyroaveraged multispecies Dougherty operator.

\subsection{LBO-G} \label{sec:lboG}

In the same vein as was done for the BGK in~\cite{Greene1973}, one may enforce exact momentum and energy conservation, and use Boltzmann relaxation rates~\citep{Morse1963} to obtain the cross-species primitive moments appropriate for Dougherty electron-ion collisions. Conservation of momentum and energy in cross-species collisions
\begin{align}
\int^\infty_{-\infty}v_i\sum_\sind \ms C[\fs]\,\text{d}^\vdim v &= 0, \label{eq:momConserv} \\
\int^\infty_{-\infty}\frac{1}{2}v^2\sum_\sind \ms C[\fs]\,\text{d}^\vdim v &= 0 \label{eq:enerConserv}
\end{align}
($C[\fs]$ the right side of equation~\ref{eq:LBDeq} with $\rind\neq\sind$) is obeyed pairwise and yields the relations
\begin{eqnal} \label{eq:conservEqs}
\sum_\sind \ms\ns\nusr\Delta\usri = 0, \\
\sum_\sind \ms\ns\nusr\left(\vdim\Delta\vtsr^2+\vus\cdot\Delta\vusr\right) = 0,
\end{eqnal}
with the sum running only over two species ($\rind$ labels the species other than $\sind$), $\Delta\vusr=\vus-\vusr$ and $\Delta\vtsr^2=\vts^2-\vtsr^2$. 
This system of $2(\vdim+1)$ unknowns can be closed in a number of ways; a particularly simple way is by employing the momentum and thermal relaxation rates of the full Coulomb collision operator (see equations 15 and 16 in~\cite{Morse1963}). For small-angle collisions these rates are
\begin{eqnal} \label{eq:rates}
\pd{}{t}\ms\ns\usi\Big|_{\mathrm{FPO}} &= \frac{\alphae}{2}\left(\ms+\mr\right)\left(\uri-\usi\right), \\
\pd{}{t}\frac{\vdim}{2}\ms\ns\vts^2\Big|_{\mathrm{FPO}} &= \frac{\alphae}{2}\left[\vdim\left(\mr\vtr^2-\ms\vts^2\right)+\mr\left(\vur-\vus\right)^2\right].
\end{eqnal}
The parameter $\alphae$ is inversely proportional to the energy and momentum relaxation times:
\begin{equation} \label{eq:alphaE}
    \alphae = \frac{2\ns\nr(\qs\qr)^2\logLambdasr}{3(2\pi)^{3/2}\epsilon_0^2\ms\mr\left(\vts^2+\vtr^2\right)^{3/2}}
\end{equation}
The right hand side of equation~\ref{eq:rates} originates from the Boltzmann collision integral for Coulomb interactions, truncated at the Debye length, under the premise that $\fs$ are close to Maxwellian. The validity of the relations given below in systems where the plasma may significantly differ from Maxwellian is thus limited.

One can compute the LBO momentum and thermal relaxation rates similar to those for the FPO in equation~\ref{eq:rates} simply by taking velocity moments of equation~\ref{eq:LBDeq}. These rates are\footnote{\cite{Dougherty1967} have an erroneous extra factor of 3 in the equivalent momentum rate of change, their equation 2.6.}
\begin{eqnal} \label{eq:ratesLBO}
\pd{}{t}\ms\ns\usi\Big|_{\mathrm{LBO}} &= \ms\ns \nusr\left(\usri-\usi\right), \\
\pd{}{t}\frac{\vdim}{2}\ms\ns\vts^2\Big|_{\mathrm{LBO}} & = \vdim\ms\ns\nusr\left(\vtsr^2-\vts^2\right).
\end{eqnal}
Equating equations~\ref{eq:rates} and~\ref{eq:ratesLBO} does not fully determine $\usr$ and $\vtsr^2$ because of the as-of-yet arbitrary $\nusr$. The next step in the Greene methodology is thus is to adopt a relationship between the collision frequency in the model operator and $\alphae$, which for the BGK operator Greene took to be $\nusr=\alphae\left(\ms+\mr\right)/[(1+\beta)\ns\ms]$ with the arbitrary parameter $\beta>-1$. For the LBO-G we will instead use
\begin{eqnal} \label{eq:nualphae}
    \nusr &= \frac{\alphae(\ms+\mr)}{\fnus(1+\beta)\ms\ns}, \\
\end{eqnal}
with $\fnus=2\mr\nr\nurs/(\ms\ns\nusr+\mr\nr\nurs)$; it turns out that $\fnus$ and $\beta$ only appear as $\fnus(1+\beta)$ so their independent values do not need to be determined separately. We picked this relationship between $\alphae$ and $\nusr$ for three reasons. First, we anticipate potential difficulties guaranteeing positivity of $\vtsr^2$, although we will see shortly that such problems do not arise with the Dougherty operator for many systems of interest. Second, the formulation presented here avoids the assumption $\ms\ns\nusr=\mr\nr\nurs$ used in earlier work~\citep{Greene1973}. Lastly, this definition of $\nusr$ produces relations that more easily enforce exact conservation in their discrete form.

Equipped with equation~\ref{eq:nualphae} we can equate equations~\ref{eq:rates} and~\ref{eq:ratesLBO}, and together with equation~\ref{eq:conservEqs} a linear system in $\usri,~\vtsr^2,~\ursi$ and $\vtrs^2$ ensues. The solution of this linear problem is
\begin{align}
\usri &= \usi + \fnus\frac{1+\beta}{2}\left(\uri-\usi\right), \label{eq:usrBetaSym} \\
\vtsr^2 &= \vts^2+\frac{\fnus}{2}\frac{1+\beta}{1+\msDmr}\left[\vtr^2-\msDmr\vts^2+\frac{\left(\vus-\vur\right)^2}{\vdim}\right], \label{eq:TsrBetaSym}
\end{align}
One attractive property of these cross-species primitive moments is that, contrary to their BGK counterparts, they do not suffer from the pathology of negative $\vtie^2$ at supersonic values of the relative drift $|\vus-\vur|$. Positivity of equation~\ref{eq:TsrBetaSym} does require however that
\begin{equation}
\frac{\fnus}{2}\frac{1+\beta}{1+\mrDms}\left[1-\TrDTs-\frac{\left(\vus-\vur\right)^2}{\vdim\cssr^2}\right] < 1,
\end{equation}
where $\cssr=\sqrt{\Ts/\mr}$. This is true for any choice of $\fnus$ and $\beta$ provided $\fnus(1+\beta)<2$, even as the relative drift increases.

Despite such improvements on previous similar multispecies operators, the unspecified $\beta$ parameter poses a clear disadvantage. \cite{Dougherty1967} had already pointed out that an additional condition is needed to determine all unknowns, and therefore avoid the appearance of any free parameters. As discussed by~\cite{Haack2017}, this free parameter can modify the transport coefficients in the associated fluid models. For the BGK operator,~\cite{Morse1964} eliminated the need for $\beta$ by assuming $\ns\nusr=\nr\nurs$ and requiring that the ratio of the relaxation rate for the momentum difference between the two species to that of the temperature difference be the same for both the FPO and the model operator. However the resulting multispecies BGK operator does not satisfy the \boltzH-theorem, discouraging us from pursuing that approach. A possible added constraint that may do away with such parameter is the isotropization rate due to interspecies collisions; imposing such condition is however beyond the scope of this manuscript.

\subsection{LBO-EM and LBO-ET}

Following the path charted by~\cite{Haack2017} for the BGK, one could require that $\usr=\urs$ and $\ms\vtsr^2=\mr\vtrs^2$. Then the momentum conservation constraint in equation~\ref{eq:conservEqs}
results in
\begin{eqnal} \label{eq:usrEq}
\usri = \frac{\ms\ns\nusr\usi+\mr\nr\nurs\uri}{\ms\ns\nusr+\mr\nr\nurs},
\end{eqnal}
while energy conservation assuming $\ms\vtsr^2=\mr\vtrs^2$ yields
\begin{eqnal} \label{eq:vtsrEq}
\left(\ns\nusr+\nr\nurs\right)\ms\vtsr^2 &= \ms\ns\nusr\vts^2+\mr\nr\nurs\vtr^2 \\
&\quad+ \frac{\ms\ns\nusr\mr\nr\nurs}{\ms\ns\nusr+\mr\nr\nurs}\frac{\left(\vus-\vur\right)^2}{\vdim}.
\end{eqnal}

The next step demands that the momentum relaxation rates are the same for both the LBO and the FPO. Setting the momentum relaxation rates equal to each other
\begin{eqnal} \label{eq:momRatesEq}
\pd{}{t}\left(\ms\ns\usi-\mr\nr\uri\right)\Big|_{\mathrm{FPO}} &= \pd{}{t}\left(\ms\ns\usi-\mr\nr\uri\right)\Big|_{\mathrm{LBO}}, \\
\alphae\left(\ms+\mr\right)\left(\uri-\usi\right) &= \ms\ns \nusr\left(\usri-\usi\right) - \mr\nr \nurs\left(\ursi-\uri\right)
\end{eqnal}
and using equation~\ref{eq:usrEq} for $\usr$ one obtains the relationship
\begin{eqnal} \label{eq:momRatesEqRes}
\alphae\left(\ms+\mr\right) &= \frac{2\ms\ns\nusr\mr\nr\nurs}{\ms\ns\nusr+\mr\nr\nurs}.
\end{eqnal}
On the other hand, equivalence between thermal relaxation rates
\begin{eqnal} \label{eq:thermalRatesEq}
\pd{}{t}\frac{\vdim}{2}\left(\ms\ns\vts^2-\mr\nr\vtr^2\right)\Big|_{\mathrm{FPO}} &= \pd{}{t}\frac{\vdim}{2}\left(\ms\ns\vts^2-\mr\nr\vtr^2\right)\Big|_{\mathrm{LBO}}, \\
\alphae\left[\vdim\left(\mr\vtr^2-\ms\vts^2\right)+\frac{\mr-\ms}{2}\left(\vus-\vur\right)^2\right] &= \vdim\left[\ms\ns\nusr\left(\vtsr^2-\vts^2\right)\right. \\
&\left.\qquad\quad-\mr\nr\nurs\left(\vtrs^2-\vtr^2\right)\right],
\end{eqnal}
with the $\vtsr^2$ from equation~\ref{eq:vtsrEq} implies that
\begin{eqnal} \label{eq:thermalRatesEqRes}
&\alphae\left[\mr\vtr^2-\ms\vts^2+\frac{\mr-\ms}{2\vdim}\left(\vus-\vur\right)^2\right] = \frac{2\ns\nusr\nr\nurs}{\ns\nusr+\nr\nurs}\left(\mr\vtr^2-\ms\vts^2\right) \\
&\quad+\frac{\ns\nusr-\nr\nurs}{\ns\nusr+\nr\nurs} \frac{\ms\ns\nusr\mr\nr\nurs}{\ms\ns\nusr+\mr\nr\nurs}\frac{\left(\vus-\vur\right)^2}{\vdim}.
\end{eqnal}
Although they may look strongly nonlinear, one can solve equations~\ref{eq:momRatesEqRes} and~\ref{eq:thermalRatesEqRes} in order to obtain an expression for $\nusr$\footnote{\cite{Haack2017} state that the equivalent equations for the BGK operator are nonlinear and without a simple formula for a solution. But one can obtain such solution by casting equation~\ref{eq:momRatesEqRes} in terms of $\tau_{\rind\sind}=1/\nurs$, solving for $\tau_{\rind\sind}$, and substituting that into equation 56 of~\cite{Haack2017} (the equivalent of our equation~\ref{eq:thermalRatesEqRes}). The result is a quadratic equation for $\ns\nusr$, which can be solved.}. The result is
\begin{equation}
\nusr = \frac{\alphae\left(\ms+\mr\right)}{\ns\ms}\cdot\frac{\frac{\ms-\mr}{2\ms\mr}\vdim\left(\mr\vtr^2-\ms\vts^2\right)+\left(\vus-\vur\right)^2}{\frac{1}{\mr}\vdim\left(\mr\vtr^2-\ms\vts^2\right)+\left(\vus-\vur\right)^2},
\end{equation}
but we can immediately notice that this can lead to negative collision frequencies in some parameter regimes; for example, the electron-ion frequency $\nuei$ with zero relative drift. This indicates that enforcing the equality of momentum and thermal relaxation rates while using the assumptions $\usri=\ursi$ and $\ms\vtsr^2=\mr\vtrs^2$ leads to unphysical behavior. We nevertheless present two slight variations in the following subsections, as was also recently done by~\cite{Ulbl2021}, in order to provide a point of reference for the LBO-G and comparing against~\cite{Haack2017}.

\subsubsection{LBO-EM} \label{sec:lboEM}

Instead of trying to match both the momentum and thermal relaxation rates, we could satisfy ourselves with only attaining the same momentum relaxation rate. We can do this by employing equation~\ref{eq:momRatesEqRes}, which we obtained from setting LBO and FPO momentum relaxation rates equal to each other, and further assuming that
\begin{equation}
\ms\ns\nusr = \mr\nr\nurs.
\end{equation}
These two equations together set the collision frequency in our model to
\begin{equation} \label{eq:nusrM}
\nusr^M = \alphae\frac{\ms+\mr}{\ms\ns} = \frac{2\left(\ms+\mr\right)(\qs\qr)^2\nr\logLambdasr}{3(2\pi)^{3/2}\epsilon_0^2\ms^2\mr\left(\vts^2+\vtr^2\right)^{3/2}}.
\end{equation}
This choice of collision frequency reduces the cross-primitive moments to
\begin{eqnal} \label{eq:crossPrimMomM}
\usri &= \frac{\usi+\uri}{2}, \\
\vtsr^2 &= \frac{1}{1+\msDmr}\left[\vts^2+\vtr^2 + \frac{\left(\vus-\vur\right)^2}{2\vdim}\right].
\end{eqnal}

We call equation~\ref{eq:LBDeq} with collision frequency and cross primitive moments in equations~\ref{eq:nusrM}-\ref{eq:crossPrimMomM} the LBO-EM. Compared to the equations that led to LBO-G, equation~\ref{eq:nusrM} suggests that LBO-EM is LBO-G in the limit of $\beta=0$ and $\fnus=1$. In this case the cross-species flow velocity in LBO-G (equation~\ref{eq:usrBetaSym}) does reduce to that in LBO-EM, but the LBO-G cross-species thermal velocity in this limit does not equal that in equation~\ref{eq:crossPrimMomM}. Interestingly for vanishing relative drifts $\beta=1$ leads to an agreement between $\vtsr^2$ for LBO-G and LBO-EM, but leads to $\usri=\uri$, which disagrees with LBO-EM's $\usri$. Therefore, as with BGK, LBO-EM is not a special case of LBO-G.

\subsubsection{LBO-ET} \label{sec:lboET}

Alternatively we could choose to approximately match the thermal relaxation rate of the FPO. Focusing on the temperature difference term in equation~\ref{eq:rates}, we see that the relaxation rate due to temperature differences alone is the same for both species. We could choose to mimic this behavior, and examining equation~\ref{eq:ratesLBO} the conclusion would be that we have to require
\begin{equation} \label{eq:nuRatioET}
    \ns\nusr = \nr\nurs.
\end{equation}
This assumption renders equations~\ref{eq:usrEq}-\ref{eq:vtsrEq} into
\begin{eqnal} \label{eq:usrEqET}
\usri = \frac{\ms\usi+\mr\uri}{\ms+\mr},
\end{eqnal}
\begin{eqnal} \label{eq:vtsrEqET}
\vtsr^2 &= \frac{1}{2}\left[\vts^2+\mrDms\vtr^2 + \frac{\mr}{\ms+\mr}\frac{\left(\vus-\vur\right)^2}{\vdim}\right].
\end{eqnal}

Although we took up relation~\ref{eq:nuRatioET} we have not specified the collision frequency precisely yet. We can do so by returning to the thermal relaxation rate equivalence (equation~\ref{eq:thermalRatesEq}) and inserting the cross-species temperature in equation~\ref{eq:vtsrEqET}. The result is\footnote{We believe there's a typo in similar equations for BGK in~\cite{Haack2017}. In equation 65 of that paper the relative drift term should be multiplied by $(m_i+3m_j)/(2m_j)$.}
\begin{eqnal}
&\alphae\left[\vdim\left(\mr\vtr^2-\ms\vts^2\right)+\frac{\mr-\ms}{2}\left(\vus-\vur\right)^2\right] = \\ &\quad\vdim\left[\frac{\ns\nusr}{2}\left(\mr\vtr^2-\ms\vts^2 + \frac{\ms\mr}{\ms+\mr}\frac{\left(\vus-\vur\right)^2}{\vdim}\right) \right.\\
&\left.\qquad- \frac{\nr\nurs}{2}\left(\ms\vts^2-\mr\vtr^2 + \frac{\ms\mr}{\ms+\mr}\frac{\left(\vus-\vur\right)^2}{\vdim}\right)\right], \\
&\alphae\left[\left(\mr\vtr^2-\ms\vts^2\right)+\frac{\mr-\ms}{2}\frac{\left(\vus-\vur\right)^2}{\vdim}\right] = \\ &\quad\frac{\ns\nusr+\nr\nurs}{2}\left(\mr\vtr^2-\ms\vts^2 \right) + \frac{1}{2}\frac{\ms\mr}{\ms+\mr}\left(\ns\nusr-\nr\nurs\right)\frac{\left(\vus-\vur\right)^2}{\vdim}.
\end{eqnal}
These thermal relaxation rates cannot agree exactly because under the assumption $\ns\nusr=\nr\nurs$ the relative drift term vanishes for this LBO, but we could at least match the response due to the temperature difference, leading to the collision frequency for this model:
\begin{equation} \label{eq:nusrT}
    \nusr^T = \frac{\alphae}{\ns} = \frac{2\nr(\qs\qr)^2\logLambdasr}{3(2\pi)^{3/2}\epsilon_0^2\ms\mr\left(\vts^2+\vtr^2\right)^{3/2}}
\end{equation}
The operator (equation~\ref{eq:LBDeq}) with this collision frequency and the cross primitive moments in equations~\ref{eq:usrEqET}-\ref{eq:vtsrEqET} is referred to as the LBO-ET. As with the previous operator, setting $\beta=0$ in the LBO-G leads to the same cross-species flow velocity $\usri$, but then the thermal speeds $\vtsr$ do not agree. More importantly, we reinstate that the LBO-ET did not exactly match the FPO thermal relaxation rate because of the difference in response to relative drifts. For plasmas where the relative drifts are small relative to temperature differences (not an uncommon situation), these rates agree exactly.

\subsection{\boltzH-theorem}

The full FPO does not decrease entropy, i.e. it satisfies an \boltzH-theorem, and as a model Fokker-Planck operator it is desirable that this formulation of multi-species Dougherty collisions retain such property. The original paper on the multispecies Dougherty operator demonstrated a non-decreasing entropy only to second order after linearization~\citep{Dougherty1967}, hinting at the possibility that the above full-$f$ equivalent operator could posses an \boltzH-theorem. It is in fact possible however to show that the Dougherty models for multi-species full-$f$ collisions presented here do have an \boltzH-theorem, even for species with unequal temperatures. A more detailed proof of this statement is given in appendix~\ref{sec:entrApp}, and we give an outline of the argument here.

The total entropy $\entr=-\sum_\sind\int\text{d}^\vdim v\thinspace \fs\ln\fs$ can be shown to obey
\begin{eqnal} \label{eq:entropyCond}
\dot{\entr} = \pd{\entr}{t} &= -\sum_\sind\int\text{d}^\vdim v~\nusr\left(\ln \fs+1\right)\divv{\vJsr},
\end{eqnal}
where the flux $\vJsr$ is the term in square brackets in equation~\ref{eq:LBDeq}. Using the definition of this flux and integration by parts twice together with the fact that $\fs\to0$ faster than powers of $v_i$ as $v_i\to\pm\infty$ one is led to
\begin{eqnal} \label{eq:sdot}
\dot{\entr} = \sum_\sind\nusr\left(-\vdim\ns+\vtsr^2\int\text{d}^\vdim v\,\gradv{\fs}\cdot\gradv{\ln\fs}\right).
\end{eqnal}
At this point we can perform a variational minimization of this functional in order to determine if that minimum is below zero (indicating a violation of thermodynamic law). For a given set of primitive moments ($\ns$, $\usi$, $\vts$, $\usri$, $\vtsr$) and the virtual displacement $\delta\fs=\fs-\fsZ$, the response of $\dot{\entr}$ is
\begin{eqnal} \label{eq:disp}
\delta\dot{\entr} = \sum_\sind\vtsr^2\int\text{d}^\vdim v\,\gradv{\fsZ}\cdot\left[\frac{2}{\delta\fs}\gradv{\delta\fs}-\frac{1}{\fsZ}\gradv{\fsZ}\right]\frac{\delta\fs}{\fsZ}.
\end{eqnal}
At an extremum in $\dot{\entr}$ this function must vanish, and since equation~\ref{eq:sdot} has no upper bound this extremum must be a minimum. We are also interested in virtual displacements that do not alter the moments of each distribution, that is
\begin{align} \label{eq:vdReq}
\int\text{d}^\vdim v~\vv^k\thinspace\delta \fs =0 \qquad \text{for } k\in\{0,1,2\}.
\end{align}
Further imposing that the displacement $\delta\fs$ vanishes at infinity, $\delta\dot{\entr}=0$ and equations~\ref{eq:disp}-\ref{eq:vdReq} yield the nonlinear inhomogeneous equation for the minimizing distribution $\fsZ$
\begin{eqnal}
\left|\gradv{\ln \fsZ}\right|^2+2\lapv{\ln \fsZ} &= h_0^2+2\vdim h_1+2h_0\v{h_1}\cdot\vv+h_1^2v^2,
\end{eqnal}
with $h_0$, $\v{h_1}$ and $h_2$ undetermined linear coefficients. The solution to this equation is $\fsZ\propto \exp\left(h_{0,i}\vi+h_1v^2/2\right)$. Enforcing the condition that it has the same number density $\ns$ and primitive moments ($\usi$ and $\vts$) as the original distribution, $\fs$, reveals that the distribution that minimizes the rate of entropy change of this operator is a Maxwellian with with $\ns$, $\usi$ and $\vts$. The final step is to insert this distribution back into our expression for entropy change, equation~\ref{eq:entropyCond}, and check that the minimum entropy rate of change does not fall below zero. Such procedure results in
\begin{eqnal} \label{eq:minS}
\min\left(\pd{\entr}{t}\right) &= \vdim \sum_\sind \frac{\ns\nusr}{\vts^2}\left(\vtsr^2-\vts^2\right),
\end{eqnal}
and since the procedure obtained a single minimum it must be the global minimum.

\subsubsection{LBO-G \boltzH-theorem}

Use the definition of $\vtsr^2$ for the LBO-G model (equation~\ref{eq:TsrBetaSym}) to arrive at
\begin{eqnal}
\min\left(\pd{\entr}{t}\right) =  \frac{\fnus\ms\ns\nusr}{2\vts^2}\frac{1+\beta}{\ms+\mr}&\left[\vdim\frac{\left(\Tr-\Ts\right)^2}{\Ts\Tr} \right.\\
&\left.\quad+\left(\frac{\mr}{\ms}\frac{1}{\vts^2}+\frac{\ms}{\mr}\frac{1}{\vtr^2}\right)\left(\vus-\vur\right)^2\right] \geq 0.
\end{eqnal}
We are thus led to the conclusion that the LBO-G model of full-$f$ multispecies collisions does not decrease the entropy. This is in contrast to the BGK-G operator, for which the \boltzH-theorem could not be proven or disproven~\citep{Greene1973,Haack2017}.

\subsubsection{LBO-EM and LBO-ET \boltzH-theorem}

Using the relationship between collision frequencies for the LBO-EM (equation~\ref{eq:nusrM}) and the corresponding cross-species thermal speeds one obtains
\begin{eqnal}
\min\left(\pd{\entr}{t}\right) = \frac{\vdim\ns\nusr}{\ms+\mr}&\left[ \frac{\left(\mr\vtr^2-\ms\vts^2\right)^2}{\vts^2\mr\vtr^2} + \left(\frac{\mr}{\vts^2}+\msDmr\frac{\ms+\mr}{\vtr^2}\right)\frac{\left(\vus-\vur\right)^2}{2\vdim}\right] \geq 0.
\end{eqnal}

Similarly, using the $\ns\nusr=\nr\nurs$ assumption of the LBO-ET model and the definition of $\vtsr^2$ in equation~\ref{eq:vtsrEqET} turns equation~\ref{eq:minS} into
\begin{eqnal}
\min\left(\pd{\entr}{t}\right) = \frac{\vdim\ns\nusr}{2}&\left[ \frac{\mr\vtr^2}{\ms\vts^2} + \frac{\ms\vts^2}{\mr\vtr^2} + \left(\frac{\mr}{\vts^2}+\frac{\ms}{\vtr^2}\right)\frac{1}{\ms+\mr}\frac{\left(\vus-\vur\right)^2}{\vdim}\right] \geq 0.
\end{eqnal}
Therefore both LBO-EM and LBO-ET models satisfy the \boltzH-theorem.

\subsection{Gyroaveraged multispecies Dougherty operator} \label{sec:GKLBO}

This model operator is also used by modern, long-wavelength full-$f$ gyrokinetic codes~\citep{Pan2018,gkeyllWeb} in its gyroaveraged form. Its form, conservative properties and discontinuous Galerkin discretization for self-species collisions have been presented by~\cite{Francisquez2020}. The operator can however be extended to incorporate cross-species collisions. For that purpose we write the gyroaveraged operator acting on the guiding center distribution function $\fs(\v{R},\vpar,\mu)$ as
\begin{eqnal} \label{eq:gkLBO}
\left(\pd{\mcJ f_\sind}{t}\right)_c = \sum_\rind\nusr &\left\{\pd{}{\vpar}\left[\left(\vpar-\uparsr\right)\mcJ\fs+\vtsr^2\pd{\mcJ\fs}{\vpar}\right] \right.\\
&\left.\quad+\pd{}{\mu}2\mu\left[\mcJ\fs+\frac{\ms\vtsr^2}{B}\pd{\mcJ\fs}{\mu}\right]\right\},
\end{eqnal}
where $\mcJ$ is the Jacobian of the guiding center coordinates, $\v{R}$ is the guiding center position, $\vpar$ is the velocity along the background magnetic field and $\mu$ is the adiabatic moment; see~\cite{Francisquez2020} for more details.

In order to use this multispecies gyroaveraged operator one must then determine the multispecies parallel flow velocities $\uparsr$ and thermal speed $\vtsr$. Our proposal is to use the same LBO-G (equations~\ref{eq:usrBetaSym},~\ref{eq:TsrBetaSym}), LBO-EM (equation~\ref{eq:crossPrimMomM}) and LBO-ET (equations~\ref{eq:usrEqET}-\ref{eq:vtsrEqET}) models with this gyroaveraged operator. The only difference is that the self-species primitive moments are defined by
\begin{eqnal}
\upars\MZs &= \MOs, \\
\upars\MOs + \vdim\vts^2\MZs &= \MTs,
\end{eqnal}
where $\vdim=1$ or $\vdim=3$ depending on whether one is considering $\vpar$- or $(\vpar,\mu)$-space, respectively. The velocity moments in the gyroaveraged model are
\begin{eqnal} \label{eq:gkMoms}
\MZs &= (2\pi/\ms)\int\mcJ\fs(\v{R},\vpar,\mu)\,\dvpar\,\dmu, \\
\MOpars &= (2\pi/\ms)\int\vpar\mcJ\fs(\v{R},\vpar,\mu)\,\dvpar\,\dmu, \\
\MTs &= (2\pi/\ms)\int\left(\vpar^2+2\mu B/\ms\right)\mcJ\fs(\v{R},\vpar,\mu)\,\dvpar\,\dmu.
\end{eqnal}

\section{Discontinuous Galerkin discretization} \label{sec:algorithm}

In this section we present a DG scheme for the multispecies LBO. DG algorithms offer higher order convergence, data locality and flexibility in defining numerical fluxes to preserve physical properties of the system~\citep{Cockburn1998,Hesthaven2007}. A DG discretization will also interface with existing Vlasov-Maxwell~\citep{Juno2018,HakimSC2020} and gyrokinetic~\citep{Shi2019,Mandell2020} DG solvers.

We present the algorithm below for a two-dimensional space consisting of one position dimension ($x$) and one velocity dimension ($v$); the extension to higher velocity dimensions is straightforward. First introduce a mesh $\mathcal{T}$ that extends over the finite computational domain $\Omega\equiv\left[-\Lx/2,\Lx/2\right]\times\left[-\Lv/2,\Lv/2\right]$ and consists of quadrilateral cells $K_{j,k}\equiv\left[x_{j-1/2},x_{j+1/2}\right]\times\left[v_{k-1/2},v_{k+1/2}\right]$, with $j=1,\dots,\Nx$ and $k=1,\dots,\Nv$ labeling the cell along $x$ and $v$, respectively. In each cell define a polynomial space $\mathcal{V}^p_{j,k}$ consisting of $\numB$ orthonormalized monomials $\psi_\ell(x,v)$, which we take as basis functions in which dynamical fields are expanded. The discretization of equation~\ref{eq:LBDeq} proceeds from a weak or Galerkin projection; multiply equation~\ref{eq:LBDeq} by $\psi_\ell$ and integrate over $x$-$v$ in cell $K_{j,k}$:
\begin{eqnal} \label{eq:dgScheme}
\int_{K_{j,k}}\psi_\ell\left(\d{\fs}{t}\right)_c\dxdv &= \int_{x_{j-1/2}}^{x_{j+1/2}}\nusr\left(\psi_\ell \Gs-\pd{\psi_\ell}{v}\vtsr^2\hat{\fs}\right)\Bigg|_{v_{k-1/2}}^{v_{k+1/2}}\dx \\
&\quad- \int_{K_{j,k}}\nusr \left[\pd{\psi_\ell}{v}\left(v-\usr\right)\fs-\pdd{\psi_\ell}{v}\vtsr^2\fs\right]\dxdv.
\end{eqnal}
We used integration by parts and limited ourselves to the case of two species cross-collisions only to remove the sum in equation~\ref{eq:LBDeq}. The numerical flux $\Gs=(v-\usr)\fs+\vtsr^2\partial \hat{\fs}/\partial v$ consists of a drag term that is computed using upwinding based on the value of the $(v-\usr)$ at Gauss-Legendre nodes, and $\hat{\fs}$ is a continuous distribution recovered across two cells~\citep{VanLeer2005,VanLeer2007}. This approach resulted in a conservative DG scheme in the case of self-species collisions; more details can be found in~\cite{Hakim2020} and~\cite{Francisquez2020}. In the case of multispecies collisions it can also lead to a conservative scheme, provided the cross-species primitive moments are computed in a manner that incorporates the finite extent of velocity-space.

In what follows we will also need the velocity moments of each species (equation~\ref{eq:moms}) in their discrete form. Discrete moments are defined as expansions in a set of $\numBx$ position-space polynomial basis functions $\varphi_\ell(x)$ belonging to the polynomial space $\mathcal{V}_k^p$ in the $j$-th cell. The discrete velocity moments are then projections of equation~\ref{eq:moms} onto the $\varphi_\ell$ basis, which for $\vdim=1$ we represent as
\begin{eqnal} \label{eq:dgMoms}
M_{q,\sind} &\doteq \int_{\vmin}^{\vmax} v^q\fs\,\dv, \qquad q\in\{0,1,2\},\\
\end{eqnal}
where $\vmin=v_{\kmin-1/2}=v_{1/2}$, $\vmax=v_{\kmax+1/2}=v_{\Nv+1/2}$, and $\doteq$ indicates weak equality~\citep{Hakim2020,Francisquez2020}. Two fields $g$ and $h$ are weakly equal in the interval $I=\left[x_{j-1/2},x_{j+1/2}\right]$ if their projections onto the basis functions in this interval are equal: $g\doteq h \Rightarrow \int_I(g-h)\varphi_\ell\,\dx=0$.

\subsection{Discrete momentum conservation}

In order to formulate a momentum conserving discretization based on equation~\ref{eq:dgScheme}, we can set $\psi_\ell=\ms v$ and sum over all cells along velocity-space. According to equation~\ref{eq:momConserv} this sum has to be equal and opposite to that of the other species it is colliding with. Therefore discrete momentum conservation requires that
\begin{eqnal}
\sum_k\int_{x_{j-1/2}}^{x_{j+1/2}}&\left(\ms\nusr\left\{\left(v \Gs-\vtsr^2\hat{\fs}\right)_{v_{k-1/2}}^{v_{k+1/2}} - \int_{v_{k-1/2}}^{v_{k+1/2}} \left(v-\usr\right)\fs\dv\right\} \right.\\
&\quad\left.+ \mr\nurs\left\{\left(v \Gr-\vtrs^2\hat{\fr}\right)_{v_{k-1/2}}^{v_{k+1/2}} - \int_{v_{k-1/2}}^{v_{k+1/2}} \left(v-\urs\right)\fr\dv\right\}\right)\dx = 0.
\end{eqnal}
Carry out the velocity-space integrals and sum over all velocity-space cells. Use the fact that the numerical fluxes $\Gs$ are continuous and have opposite signs on either side of a cell boundary, and that $\hat{\fs}$ is continuous across cell boundaries as well. Furthermore, we impose the zero-flux boundary conditions $\Gs(v=\vmax)=\Gs(v=\vmin)=0$ in order to arrive at
\begin{eqnal}
\int_{x_{j-1/2}}^{x_{j+1/2}}&\left[\ms\nusr\left(\vtsr^2\fs\Big|_{\vmin}^{\vmax} + \MOs - \usr\MZs\right) \right.\\
&\quad\left.+ \mr\nurs\left(\vtrs^2\fr\Big|_{\vmin}^{\vmax} + \MOr - \urs\MZr\right)\right]\dx = 0,
\end{eqnal}
having substituted the discrete form of the velocity moments (equation~\ref{eq:dgMoms}). This relation is satisfied if
\begin{eqnal} \label{eq:momC}
\ms\nusr\left(\vtsr^2\fs\Big|_{\vmin}^{\vmax} + \MOs - \usr\MZs\right) + \mr\nurs\left(\vtrs^2\fr\Big|_{\vmin}^{\vmax} + \MOr - \urs\MZr\right) \doteq 0.
\end{eqnal}
Note that we have used the same $\vmax$ and $\vmin$ for both species for pedagogical reasons only. In fact since equation~\ref{eq:momC} is a constraint on position-space fields only, the discrete velocity-space of each species can be completely different. For completeness we state that in the $\vdim$-dimensional case the condition for the scheme in equation~\ref{eq:dgScheme} to conserve momentum is that the cross-species primitive moments $\usr$ and $\vtsr$ must satisfy
\begin{eqnal} \label{eq:momCvdim}
&\ms\nusr\left(\MOis - \usri\MZs + \vtsr^2\int\fs\Big|_{\vimin}^{\vimax}\mathrm{d}S_i\right) \\
&\quad+ \mr\nurs\left(\MOir - \ursi\MZr + \vtrs^2\int\fr\Big|_{\vimin}^{\vimax}\mathrm{d}S_i\right) \doteq 0,
\end{eqnal}
using $\int\mathrm{d}S_i$ as an integral over the velocity-space boundaries orthogonal to the $i$-th velocity dimension, and the repeated index $i$ implies summation.

\subsection{Discrete energy conservation} \label{sec:discreteEnerConserv}

Energy conservation will impose a secondary constraint on how the discrete cross-species primitive moments must be computed. In order to obtain such condition we substitute $\psi_\ell=\ms v^2/2$ in equation~\ref{eq:dgScheme}, and sum over velocity-space cells and species. This action leads to
\begin{eqnal}
&\sum_k\int_{x_{j-1/2}}^{x_{j+1/2}}\left(\ms\nusr\left\{\left(\frac{v^2}{2}\Gs-v\vtsr^2\hat{\fs}\right)\Big|_{v_{k-1/2}}^{v_{k+1/2}} - \int_{v_{k-1/2}}^{v_{k+1/2}} \left[v\left(v-\usr\right)\fs-\vtsr^2\fs\right]\dv\right\} \right.\\
&\left.+\mr\nurs\left\{\left(\frac{v^2}{2}\Gr-v\vtrs^2\hat{\fr}\right)\Big|_{v_{k-1/2}}^{v_{k+1/2}} - \int_{v_{k-1/2}}^{v_{k+1/2}} \left[v\left(v-\urs\right)\fr-\vtrs^2\fr\right]\dv\right\}\right)\dx = 0.
\end{eqnal}
Once again we employ continuity of $\Gs$ and $\hat{\fs}$ and boundary conditions so that after performing the velocity-space integrals and carrying out the sum over velocity-space cells this relation is transformed into
\begin{eqnal}
&\int_{x_{j-1/2}}^{x_{j+1/2}}\left(\ms\nusr\left\{v\vtsr^2\fs\Big|_{\vmin}^{\vmax} + \left(\MTs-\usr\MOs\right)-\vtsr^2\MZs\right\} \right.\\
&\left.\quad+\mr\nurs\left\{v\vtrs^2\fr\Big|_{\vmin}^{\vmax} + \left(\MTr-\urs\MOr\right)-\vtrs^2\MZr\right\}\right)\dx = 0.
\end{eqnal}
Therefore we can guarantee that our DG discretization exactly conserves energy if we enforce
\begin{eqnal} \label{eq:enerC}
&\ms\nusr\left\{v\vtsr^2\fs\Big|_{\vmin}^{\vmax} + \left(\MTs-\usr\MOs\right)-\vtsr^2\MZs\right\} \\
&\quad+\mr\nurs\left\{v\vtrs^2\fr\Big|_{\vmin}^{\vmax} + \left(\MTr-\urs\MOr\right)-\vtrs^2\MZr\right\} \doteq 0
\end{eqnal}
when computing $\usr$, $\urs$, $\vtsr$ and $\vtrs$. In the case of $\vdim$ velocity dimensions this constraint becomes
\begin{eqnal} \label{eq:enerCvdim}
&\ms\nusr\left[\MTs-\usri\MOis-\vtsr^2\left(\vdim\MZs-\int v_i\fs\Big|_{\vimin}^{\vimax}\mathrm{d}S_i\right)\right] \\
&\quad+\mr\nurs\left[\MTr-\ursi\MOir-\vtrs^2\left(\vdim\MZr-\int v_i\fr\Big|_{\vimin}^{\vimax}\mathrm{d}S_i\right)\right] \doteq 0.
\end{eqnal}
We make a final comment that the substitution $\psi_\ell=\ms v^2/2$ is only valid if $v^2$ belongs to the space spanned by the basis, which for piecewise linear basis functions ($p=1$) it does not. In order for the algorithm to be conservative with $p=1$ additional precautions must be taken, a topic that is deferred to appendix~\ref{sec:pOenerConserv}.

\subsection{Discrete relaxation rates}

Together with the relations $\usri=\ursi$ and $\ms\vtsr^2=\mr\vtrs^2$, and the definitions of the collision frequency given in sections~\ref{sec:lboEM}-\ref{sec:lboET}, equations~\ref{eq:momCvdim} and~\ref{eq:enerCvdim} are all one needs to compute the cross-primitive moments for the LBO-EM and LBO-ET. The LBO-G however needs to further incorporate the equivalence between the momentum and thermal relaxation rates of the FPO (equation~\ref{eq:rates}) and those of the LBO (equations~\ref{eq:ratesLBO}) in the discrete sense. First we obtain the weak form of equation~\ref{eq:momRatesEq} by projecting it onto the $\psi_\ell$ basis function:
\begin{eqnal} \label{eq:discreteMomRateMatch}
&\ms\nusr\left(\MOis - \usri\MZs + \vtsr^2\int\fs\Big|_{\vimin}^{\vimax}\mathrm{d}S_i\right) \\
&\quad- \mr\nurs\left(\MOir - \ursi\MZr + \vtrs^2\int\fr\Big|_{\vimin}^{\vimax}\mathrm{d}S_i\right) \doteq \alphae\left(\ms+\mr\right)\left(\usi-\uri\right),
\end{eqnal}
where we used the same series of steps that led to equation~\ref{eq:momCvdim}.

The equivalent condition on the thermal relaxation rates necessitates the discrete thermal speed moment of the LBO. We can obtain it by substituting $\psi_\ell=\ms\left(v-\us\right)^2/2$ into the weak scheme in equation~\ref{eq:dgScheme} and summing over velocity-space cells, resulting in
\begin{eqnal} \label{eq:discThermalRelaxRate}
&\sum_k\int_{K_{j,k}}\frac{\ms}{2}\left(v-\us\right)^2\left(\d{\fs}{t}\right)_c\dxdv = \sum_k\int_{x_{j-1/2}}^{x_{j+1/2}}\ms\nusr\left\{\left[\frac{1}{2}\left(v-\us\right)^2\Gs \right.\right.\\
&\left.\left.\quad- \left(v-\us\right)\vtsr^2\hat{\fs}\right]\Big|_{v_{k-1/2}}^{v_{k+1/2}} - \int_{v_{k-1/2}}^{v_{k+1/2}} \left[\left(v-\us\right)\left(v-\usr\right)-\vtsr^2\right]\fs\dv\right\}\dx.
\end{eqnal}
Performing the velocity-space integrals, carrying out the $k$-sum, accounting for the continuity of $\Gs$ and $\hat{\fs}$ and using the zero-flux boundary conditions on $\Gs$ ushers us to
\begin{eqnal}
&\sum_k\int_{K_{j,k}}\left(\d{\left(\ms\ns\vts^2/2\right)}{t}\right)_c\dxdv = \int_{x_{j-1/2}}^{x_{j+1/2}}\ms\nusr\left\{\left[- \left(v-\us\right)\vtsr^2\fs\right]\Big|_{\vmin}^{\vmax} \right.\\
&\left.\quad- \left[\MTs-\us\MOs+\usr\left(\us\MZs-\MOs\right)-\vtsr^2\MZs\right]\right\}\dx,
\end{eqnal}
or in the $\vdim$-dimensional velocity space
\begin{eqnal}
&\sum_k\int_{K_{j,k}}\left(\d{\left(\ms\ns\vts^2/2\right)}{t}\right)_c\dxdv = \int_{x_{j-1/2}}^{x_{j+1/2}}\ms\nusr\left\{\vtsr^2\int\left[- \left(v_i-\usi\right)\fs\right]\Big|_{\vimin}^{\vimax}\mathrm{d}S_i \right.\\
&\left.\quad- \left[\MTs-\usi\MOis+\usri\left(\usi\MZs-\MOis\right)-\vtsr^2\vdim\MZs\right]\right\}\dx,
\end{eqnal}
where once again the repeated index $i$ implies summation. Equipped with this formula we can write down the discrete equivalence between thermal relaxation rates (equation~\ref{eq:thermalRatesEq}) as
\begin{eqnal} \label{eq:discreteThermalRateMatch}
&\ms\nusr\left\{\vtsr^2\int\left[- \left(v_i-\usi\right)\fs\right]\Big|_{\vimin}^{\vimax}\mathrm{d}S_i \right.\\
&\left.\quad- \left[\MTs-\usi\MOis+\usri\left(\usi\MZs-\MOis\right)-\vtsr^2\vdim\MZs\right]\right\} \\
&\quad- \mr\nurs\left\{\vtrs^2\int\left[- \left(v_i-\uri\right)\fr\right]\Big|_{\vimin}^{\vimax}\mathrm{d}S_i \right.\\
&\left.\quad- \left[\MTr-\uri\MOir+\ursi\left(\uri\MZr-\MOir\right)-\vtrs^2\vdim\MZr\right]\right\} \\
&\quad\doteq \alphae \left[\vdim\left(\mr\vtr^2-\ms\vts^2\right)+\frac{\mr-\ms}{2}\left(\vus-\vur\right)^2\right]
\end{eqnal}
The two discrete relaxation rate equivalences in equations~\ref{eq:discreteMomRateMatch} and~\ref{eq:discreteThermalRateMatch} in conjunction with the discrete momentum and energy conservation constraints (equations~\ref{eq:momCvdim} and~\ref{eq:enerCvdim}) provide the four equations for the calculation of the $\usri,~\ursi,~\vtsr,~\vtrs$ unknowns in the LBO-G, provided a value of $\beta$. Equations~\ref{eq:discreteMomRateMatch} and~\ref{eq:discreteThermalRateMatch} however are written in terms of the self primitive moments (e.g. $\usi,~\vts$) of each species, imbuing such equations with some ambiguity as to whether the calculation of self-primitive moments should include the corrections from velocity-space boundaries or not~\citep{Hakim2020,Francisquez2020}. We therefore opt to instead write those relations in terms of the velocity moments as follows
\begin{eqnal}
&\ms\nusr\left(\MOis - \usri\MZs + \vtsr^2\int\fs\Big|_{\vimin}^{\vimax}\mathrm{d}S_i\right) \\
&\quad- \mr\nurs\left(\MOir - \ursi\MZr + \vtrs^2\int\fr\Big|_{\vimin}^{\vimax}\mathrm{d}S_i\right) \\
&\quad\doteq \frac{\alphae\left(\ms+\mr\right)}{\MZs\MZr}\left(\MZr\MOis-\MZs\MOir\right), \\
&\ms\nusr\left\{-\vtsr^2\int \left(v_i-\usi\right)\fs\Big|_{\vimin}^{\vimax}\mathrm{d}S_i \right.\\
&\left.\quad- \left[\MTs-\usi\MOis+\usri\left(\usi\MZs-\MOis\right)-\vtsr^2\vdim\MZs\right]\right\} \\
&\quad- \mr\nurs\left\{-\vtrs^2\int \left(v_i-\uri\right)\fr\Big|_{\vimin}^{\vimax}\mathrm{d}S_i \right.\\
&\left.\quad- \left[\MTr-\uri\MOir+\ursi\left(\uri\MZr-\MOir\right)-\vtrs^2\vdim\MZr\right]\right\} \\
&\quad\doteq \frac{\alphae}{\MZs\MZr}\left[\mr\MZs\left(\MTr-\uri\MOir\right)-\ms\MZr\left(\MTs-\usi\MOis\right) \right.\\
&\left.\quad+\frac{\mr-\ms}{2}\left(\usi-\uri\right)\left(\MZr\MOis-\MZs\MOir\right)\right].
\end{eqnal}
The division by $\MZs\MZr$ on the right side of these equations is to be performed weakly~\citep{Hakim2020} in order to avoid aliasing errors.

\subsection{Summary of discrete equations}

In summary, in order for the algorithm based on equation~\ref{eq:dgScheme} to conserve momentum and energy the discrete cross primitive moments are computed using the conservation constraints
\begin{eqnal} \label{eq:LBOEeqns}
&\ms\nusr\left(\usri\MZs - \vtsr^2\int\fs\Big|_{\vimin}^{\vimax}\mathrm{d}S_i\right) \\
&\quad+ \mr\nurs\left(\ursi\MZr - \vtrs^2\int\fr\Big|_{\vimin}^{\vimax}\mathrm{d}S_i\right) \doteq \ms\nusr\MOis + \mr\nurs\MOir, \\
&\ms\nusr\left[\usri\MOis + \vtsr^2\left(\vdim\MZs-\int v_i\fs\Big|_{\vimin}^{\vimax}\mathrm{d}S_i\right)\right] \\
&\quad+\mr\nurs\left[\ursi\MOir + \vtrs^2\left(\vdim\MZr-\int v_i\fr\Big|_{\vimin}^{\vimax}\mathrm{d}S_i\right)\right] \doteq \ms\nusr\MTs + \mr\nurs\MTr.
\end{eqnal}
Additionally, the LBO-EM and LBO-ET use $\usri=\ursi$ and $\ms\vtsr^2=\mr\vtrs^2$, respectively, as well as their corresponding collision frequencies (equations~\ref{eq:nusrM} and~\ref{eq:nusrT}). When the discrete expansions are inserted in equation~\ref{eq:LBOEeqns} one is faced with a linear problem of size $(\vdim+1)\numBx$ that must be solved in every position-space cell ($\numBx$ is the number of monomials of the basis spanning position-space). The LBO-G on the other hand uses the equality between discrete LBO relaxation rates and the FPO relaxation rates:
\begin{eqnal} \label{eq:discreteMomRelax}
&\ms\nusr\left(\usri\MZs - \vtsr^2\int\fs\Big|_{\vimin}^{\vimax}\mathrm{d}S_i\right) - \mr\nurs\left(\ursi\MZr - \vtrs^2\int\fr\Big|_{\vimin}^{\vimax}\mathrm{d}S_i\right) \\
&\quad\doteq \ms\nusr\MOis - \mr\nurs\MOir + \frac{\alphae\left(\ms+\mr\right)}{\MZs\MZr}\left(\MZs\MOir-\MZr\MOis\right),
\end{eqnal}
\begin{eqnal} \label{eq:discreteThermalRelax}
&\ms\nusr\left\{\usri\left(\MOis-\usi\MZs\right)+\vtsr^2\left[\vdim\MZs - \int\left(v_i-\usi\right)\fs\Big|_{\vimin}^{\vimax}\mathrm{d}S_i\right]\right\} \\
&\quad- \mr\nurs\left\{\ursi\left(\MOir-\uri\MZr\right)+\vtrs^2\left[\vdim\MZr - \int\left(v_i-\uri\right)\fr\Big|_{\vimin}^{\vimax}\mathrm{d}S_i\right]\right\} \\
&\quad\doteq \ms\nusr\left(\MTs-\usi\MOis\right) - \mr\nurs\left(\MTr-\uri\MOir\right) \\
&\quad+ \frac{\alphae}{\MZs\MZr}\left[\mr\MZs\left(\MTr-\uri\MOir\right)-\ms\MZr\left(\MTs-\usi\MOis\right) \right.\\
&\left.\quad+\frac{\mr-\ms}{2}\left(\usi-\uri\right)\left(\MZr\MOis-\MZs\MOir\right)\right],
\end{eqnal}
where the relationship between $\alphae$ and $\nusr$ is given by equation~\ref{eq:nualphae}. Therefore for the LBO-G equations~\ref{eq:LBOEeqns}-\ref{eq:discreteThermalRelax} signify a $2(\vdim+1)\numBx$ linear problem that must be solved in every position-space cell.

\subsubsection{Discrete equations for the gyroaveraged operator} \label{sec:discreteGkLBO}

In section~\ref{sec:GKLBO} we introduced the gyroaveraged cross-species LBO. Its discretization follows that outlined in~\cite{Francisquez2020} and the calculation of the cross primitive moments $\uparsr$ and $\vtsr$ is similar to that done for the non-gyroaveraged operator. The main differences arise from the fact that moments are defined via $\vpar$-$\mu$ integrals (e.g. equation~\ref{eq:gkMoms}) and that the momentum density $\MOpars$ is a scalar instead of a vectorial quantity. The equations that arise from momentum and energy conservation in the gyroaveraged case are thus
\begin{eqnal} \label{eq:momCvdimGK}
&\ms\nusr\left(\MOpars - \uparsr\MZs + \vtsr^2\frac{2\pi}{\ms}\int\mcJ\hat{\fs}\Big|_{\vparmin}^{\vparmax}\dmu\right) \\
+&\mr\nurs\left(\MOparr - \uparrs\MZr + \vtrs^2\frac{2\pi}{\mr}\int\mcJ\hat{\fr}\Big|_{\vparmin}^{\vparmax}\dmu\right) \doteq 0,
\end{eqnal}
\begin{eqnal} \label{eq:enerCvdimGK}
&\ms\nusr\left[\MTs-\uparsr\MOpars\right.\\
&\left.\quad-\vtsr^2\left(3\MZs-\frac{2\pi}{\ms}\int \vpar\mcJ\hat{\fs}\Big|_{\vparmin}^{\vparmax}\dmu-\frac{2\pi}{\ms}\int 2\mu\mcJ\hat{\fs}\Big|_{\mumin}^{\mumax}\dvpar\right)\right] \\
+&\mr\nurs\left[\MTr-\uparrs\MOparr \right.\\
&\left.\quad-\vtrs^2\left(3\MZr-\frac{2\pi}{\mr}\int\vpar\mcJ\hat{\fr}\Big|_{\vparmin}^{\vparmax}\dmu-\frac{2\pi}{\mr}\int 2\mu\mcJ\hat{\fr}\Big|_{\mumin}^{\mumax}\dvpar\right)\right]\doteq 0,
\end{eqnal}
For the gyroaveraged LBO-EM and LBO-ET equations~\ref{eq:momCvdimGK}-\ref{eq:enerCvdimGK}, and the relations $\uparsr=\uparrs$ and $\ms\vtsr^2=\mr\vtrs^2$, is all that is needed to compute the cross primitive moments. This requires a solution to a linear problem of size $2\numBx$ in each position-space cell. In the case of the LBO-G operator we make use of the discrete moment relaxation equations once again, as in equations~\ref{eq:discreteMomRelax}-\ref{eq:discreteThermalRelax} but using the gyroaveraged moments instead:
\begin{eqnal} \label{eq:discreteMomRelaxGK}
&\ms\nusr\left(\uparsr\MZs - \vtsr^2\frac{2\pi}{\ms}\int\mcJ\hat{\fs}\Big|_{\vparmin}^{\vparmax}\dmu\right) - \mr\nurs\left(\uparrs\MZr - \vtrs^2\frac{2\pi}{\mr}\int\mcJ\hat{\fr}\Big|_{\vparmin}^{\vparmax}\dmu\right) \\
&\quad\doteq \ms\nusr\MOpars - \mr\nurs\MOparr + \frac{\alphae\left(\ms+\mr\right)}{\MZs\MZr}\left(\MZs\MOparr-\MZr\MOpars\right),
\end{eqnal}
\begin{eqnal} \label{eq:discreteThermalRelaxGK}
&\ms\nusr\left\{\uparsr\left(\MOpars-\upars\MZs\right) \right.\\
&\left.\quad+\vtsr^2\left[3\MZs - \frac{2\pi}{\ms}\int\left(\vpar-\upars\right)\mcJ\hat{\fs}\Big|_{\vparmin}^{\vparmax}\dmu - \frac{2\pi}{\ms}\int2\mu\mcJ\hat{\fs}\Big|_{\mumin}^{\mumax}\dvpar\right]\right\} \\
-&\mr\nurs\left\{\uparrs\left(\MOparr-\uparr\MZr\right) \right.\\
&\left.\quad+\vtrs^2\left[3\MZr - \frac{2\pi}{\mr}\int\left(\vpar-\uparr\right)\mcJ\hat{\fr}\Big|_{\vparmin}^{\vparmax}\dmu - \frac{2\pi}{\mr}\int2\mu\mcJ\hat{\fr}\Big|_{\mumin}^{\mumax}\dvpar\right]\right\} \\
\quad&\doteq \ms\nusr\left(\MTs-\upars\MOpars\right) - \mr\nurs\left(\MTr-\uparr\MOparr\right) \\
&\quad+ \frac{\alphae}{\MZs\MZr}\left[\mr\MZs\left(\MTr-\uparr\MOparr\right)-\ms\MZr\left(\MTs-\upars\MOpars\right) \right.\\
&\left.\quad+\frac{\mr-\ms}{2}\left(\upars-\uparr\right)\left(\MZr\MOpars-\MZs\MOparr\right)\right],
\end{eqnal}
where once again weak division by $\MZs\MZr$ on the right side of these equations assumed~\citep{Hakim2020} in order to avoid aliasing errors. In equations~\ref{eq:momCvdimGK}-\ref{eq:discreteThermalRelaxGK} we assumed a $\vpar$-$\mu$ simulation such that $\vdim=3$. For a $\vpar$ simulation the 3 in front of $\MZs$ and $\MZr$ would be simply a 1, $(2\pi/m)\int\dmu$ integrals would vanish and so would the the $2\mu\mcJ\hat{f}$ terms. In either case the gyroaveraged LBO-G requires inverting a matrix with $4\numBx\times4\numBx$ matrix in each configuration space cell.

\section{Benchmarks and results} \label{sec:results}

The algorithm introduced in section~\ref{sec:algorithm} has been implemented in the DG Vlasov-Maxwell~\citep{Juno2018,HakimSC2020} and gyrokinetic~\citep{Shi2019,Mandell2020} solvers of the \gkeyll~computational plasma physics framework~\citep{gkeyllWeb}. In order to demonstrate the algorithm's properties and test the implementation we have run a number of tests and we here present the results of three of them: section~\ref{sec:conservTest} contains basic tests showing the conservative properties of the algorithm, section~\ref{sec:landauTest} contains four dimensional Vlasov-Maxwell simulations of collisional Landau damping of an electron plasma wave, and section~\ref{sec:relaxTest} uses the gyrokinetic solver to explore velocity and temperature relaxation. All the input files used to generate these results are available online (see appendix~\ref{sec:getGkeyll}).

\subsection{Conservation tests} \label{sec:conservTest}

\subsubsection{Vlasov LBO conservation} \label{sec:vmConservTest}

We check that momentum and energy are indeed conserved by our discrete scheme by initiating two populations of electrons and protons with an arbitrary non-Maxwellian distribution function given by:
\begin{equation} \label{eq:vmConservIC}
    f_\sind(\vv,t=0) = a\left[1+d\cos\left(k_\sind v\right)\right]\exp\left[-\frac{\left(\vv-\v{b}_\sind\right)^2}{2\sigma_\sind^2}\right],
\end{equation}
with $a=7\times10^{19}$, $d=0.5$, $k_\sind=\pi/\vts$, $\sigma_\sind=\vts$, $\v{b}_e=\left\{\vte/2,-\vte/2,0\right\}$, $\v{b}_i=\left\{3\vti/2,3\vti/2,\vti/2\right\}$ with $\vts=\sqrt{T_{\sind0}/\ms}$, and $T_{e0}=40$ eV and $T_{i0}=80$ eV. We discretize these distribution functions in phase spaces restricted to $\left[-1,1\right]\times\left[-5\vts,5\vts\right]^{\vdim}$ and meshed with $1\times\Nv^{\vdim}$ cells. We show conservation properties for both piecewise linear ($p=1$) and piecewise quadratic ($p=2$) Serendipity basis functions~\citep{Juno2018}; higher order basis functions may also be used but the results do not change. For the same reason we use a single cell in position-space; the results in this section are independent of position-space dimensionality although we checked such cases anyway to make sure there are no errors in the implementation.

\begin{figure}
  \centering
  \includegraphics[width=0.65\textwidth]{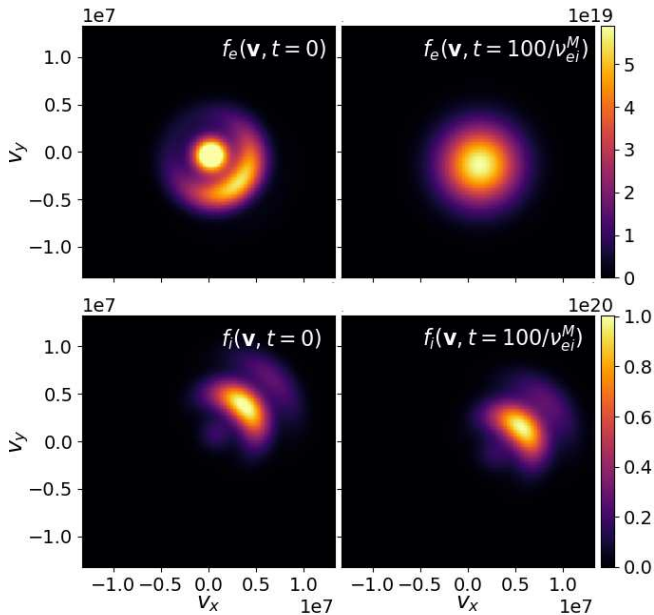}
  \caption{Initial and final (after $\nuei^M t=100$) electron and ion distribution functions as they collide with each other with the LBO-EM model using $1\times64^2$ cells and a $p=2$ Serendipity basis. Initial conditions are given in equation~\ref{eq:vmConservIC}. Colorbars are normalized to the extrema at $\nuei^M t=100$.}
  \label{fig:vmConservDistf}
\end{figure}

We time-integrate the cross-species collision terms (no self-species collisional or collisionless terms are included here) with constant collision frequency using a strong-stability preserving (SSP) third-order Runge-Kutta method (RK3). As electrons and ions collide with each other their temperatures and flow velocities relax to a common value, a process that is more carefully benchmarked in section~\ref{sec:relaxTest}. We also see that whatever anisotropies were present at $t=0$ go away on the $\nusr^{-1}$ time scale. In figure~\ref{fig:vmConservDistf}, for example, we illustrate the isotropization of the electrons after several $(\nuei^M)^{-1}$ periods as they collide with the ions using the LBO-EM, but since $\nuie$ is smaller by $m_e/m_i$ the ions will take much longer to isotropize as they collide with the electrons.

\begin{figure}
  \centering
  \includegraphics[width=0.75\textwidth]{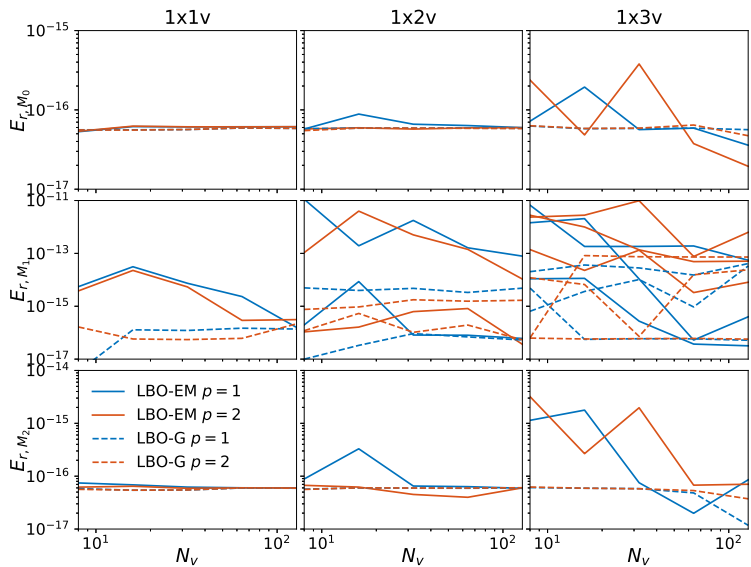}
  \caption{Relative error per time step in number density (top row), momentum density in each direction (middle row) and kinetic energy density (bottom row). The middle row plots contain $\vdim$ lines for each operator and $p$, corresponding to the error per time step in the conservation of momentum along each direction.}
  \label{fig:vmConserv}
\end{figure}

We ran this simulation for $\vdim=\{1,2,3\}$, $p=\{1,2\}$ and using both the LBO-EM and the LBO-G. The ability to conserve the first three volume-integrated velocity moments of the distribution function was quantified in each case by integrating the equations for $N_t$ time steps, and computing the relative error per time step in the volume integrated particle, momentum and kinetic energy density. The relative error per time step in the number density $M_0$ is given by
\begin{equation}
    E_{r,M_0} = \frac{1}{N_t}\frac{\avg{M_{0e}+M_{0i}}(t=N_t\Delta t)-\avg{M_{0e}+M_{0i}}(t=0)}{\avg{M_{0e}+M_{0i}}(t=0)},
\end{equation}
where $\avg{\cdot}$ indicates a volume average and $N_t=10^4$. The relative error per time step in momentum and kinetic energy conservation take into account the mass of each species:
\begin{align}
    E_{r,M_{1k}} = \frac{1}{N_t}\frac{\avg{m_eM_{1e,k}+m_iM_{1i,k}}(t=N_t\Delta t)-\avg{m_eM_{1e,k}+m_iM_{1i,k}}(t=0)}{\avg{m_eM_{1e,k}+m_iM_{1i,k}}(t=0)}, \\
    E_{r,M_2} = \frac{1}{N_t}\frac{\avg{\frac{1}{2}m_eM_{2e}+\frac{1}{2}m_iM_{2i}}(t=N_t\Delta t)-\avg{\frac{1}{2}m_eM_{2e}+\frac{1}{2}m_iM_{2i}}(t=0)}{\avg{\frac{1}{2}m_eM_{2e}+\frac{1}{2}m_iM_{2i}}(t=0)}.
\end{align}
The results as a function of velocity space resolution (i.e. $N_v$) are given in figure~\ref{fig:vmConserv}. The middle row plots have $\vdim$ lines for each operator and polynomial order $p$ because the relative error per time step in the volume-integrated momentum density is measured along each direction separately. In all cases we see that the errors in momentum conservation per time step remain of the order of machine precision. This is true even for the simulations with piece-wise linear basis functions or very coarse velocity-space meshes. The LBO-ET uses the same algorithm and implementation as LBO-EM but with a different collision frequency, so its conservation errors are similar to those of the LBO-EM shown here.

\begin{figure}
  \centering
  \includegraphics[width=0.8\textwidth]{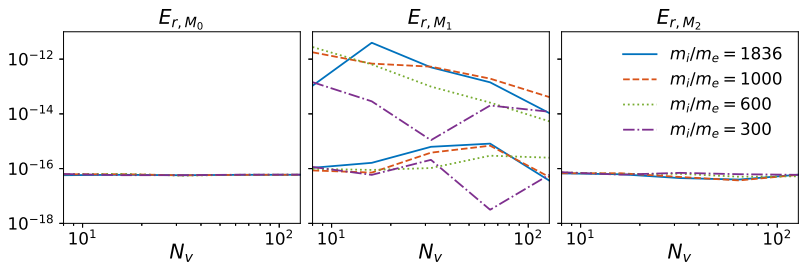}
  \caption{Relative error per time step in number density (left), momentum density in each direction (center) and kinetic energy density (right) in two-dimensional velocity space ($\vdim=2$).}
  \label{fig:vmConservMassRatScan}
\end{figure}

These conservation properties do not depend on the large mass disparity between ions and electrons; the algorithm's ability to conserve the velocity moments is also independent of the mass ratio. We provide as an example the $\vdim=2$ and $p=2$ simulation with the LBO-EM, scanning the number of velocity space cells in one direction ($\Nv$) and using the mass ratios $m_i/m_e=\{300,600,1000,1836\}$. The conservation errors for these simulations are provided in figure~\ref{fig:vmConservMassRatScan}, once again shown that for all mass ratios and resolutions used, the error per time step in the volume integrated velocity moments remains of the order of machine precision.

\subsubsection{Gyroaveraged LBO conservation} \label{sec:gkConservTest}

Similar tests were run with the gyroaveraged version of the LBO operators in order to guarantee that the algorithm remains conservative in that case as well. For these tests we initialize the ion and electron distribution functions with
\begin{equation} \label{eq:gkConservIC}
    f_\sind(\vpar,\mu,t=0) = \frac{a\left[1+d\cos\left(k_\sind v\right)\right]}{\left(2\pi\sigma_\sind^2\right)^{3/2}}\exp\left[-\frac{\left(\vpar-b_\sind\right)^2+2\mu B\ms}{2\sigma_\sind^2}\right],
\end{equation}
and the parameters $B=1.2$ T, $a=7\times10^{19}$, $d=0.5$, $k_\sind=\pi/\vts$, $\sigma_\sind=\vts$, $b_e=5\vti/4$, $b_i=\vti$, $\vts=\sqrt{T_{\sind 0}/\ms}$, $T_{e0}=40$ eV and $T_{i0}=80$ eV. The phase space $[-1,1]\times\left[-5\vts,5\vts\right]\times\left[0,\ms\left(5\vts\right)^2/(2B)\right]$ is meshed with $1\times\Nv^2$ cells and functions are expanded on piecewise linear ($p=1$) or piecewise quadratic ($p=2$) Serendipity basis.

\begin{figure}
  \centering
  \includegraphics[width=0.75\textwidth]{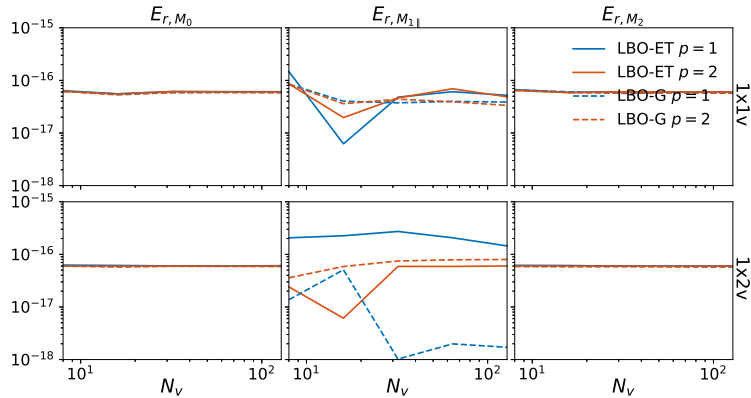}
  \caption{Relative error per time step in number density (left), momentum density (center) and kinetic energy density (right) as a function of the number of cells along one direction of velocity space ($\Nv$) for the gyroaveraged LBO-ET (solid) and LBO-G (dashed). Top row contains tests with $\vpar$-space only, while the bottom row contains tests in $\vpar-\mu$ space.}
  \label{fig:gkConserv}
\end{figure}

We allow the electrons and ions to collide with each other but not with themselves, and we do not apply the collisionless terms either. The cross-species collision terms were integrated in time for $10^4$ time steps using a third-order SSP RK3, and we computed the relative error per time step in the volume integrated velocity moments as in section~\ref{sec:gkConservTest}. The results in figure~\ref{fig:gkConserv} demonstrate how the relative error per time step in the conservation of velocity moments stays of order of machine precision for all velocity-space resolutions, and even for $p=1$. Figure~\ref{fig:gkConserv} gives conservation errors for the LBO-ET and the LBO-G; the LBO-EM has similar conservative properties as the LBO-ET since it only differs by the definition of the collision frequency.

\subsection{Landau damping of electron Langmuir waves} \label{sec:landauTest}

A seminal test-bed for collision operators is the Landau damping of plasma waves
across the collisional range. We pursued this analysis to examine the effect that these collision models have on the Landau damping rate of electrostatic electron Langmuir waves. For this purpose we employ the Vlasov-Maxwell solver in \gkeyll~\citep{Juno2018,HakimSC2020} with the self-species collision terms~\citep{Hakim2020} and the multi-species collision models described in this work. The hydrogen ions are fixed in time so the equations solved are
\begin{align}
    \pd{f_e}{t} + \vv\cdot\grad{f_e} - \v{E}\cdot\gradv{f_e} &
    = \sum_{\rind=e,i}\nu_{e\rind} \divv{\left[\left(\vv-\vu_{e\rind}\right)f_e+v_{te\rind}^2\gradv{f_e}\right]}, \\
    \pd{\v{E}}{t} &= -\v{J}, \label{eq:AmperesEq}
\end{align}
where we used normalized units\footnote{https://gkeyll.readthedocs.io/en/latest/dev/vlasov-normalizations.html}, the current density is given by $\v{J}=-\v{M_{1e}}$, and we solve equation~\ref{eq:AmperesEq} in a way that keeps the simulation electrostatic~\footnote{http://ammar-hakim.org/sj/je/je33/je33-buneman.html}. We use four-dimensional simulations with the phase-space $\left[-\pi/k,\pi/k\right]\times\left[-5\vte,5\vte\right]^3$ discretized by $16\times36^3$ cells and $p=2$ basis functions, for the wavenumber $k\lambda_{De}=0.3$, with $\lambda_{De}$ being the electron Debye length. We confirmed that the resolution used is the minimum needed to obtain converged results by scanning the position- and velocity-space resolution as well as the velocity-space extents. The static ions have the normalized density $n_i(x)=1$ while the electrons are initialized with a non-drifting Maxwellian distribution that has the temperature $T_e=1=T_i$ and the density $n_e(x,t=0)=1+\alpha\cos\left(kx\right)$, with $\alpha=10^{-4}$. The electric field is initialized in a manner consistent with Poisson's equation: $\v{E}=-\uv{x}\, \alpha\sin(kx)/k$. 

\begin{figure}
  \centering
  \begin{subfigure}[b]{0.495\textwidth}
    \includegraphics[width=\textwidth]{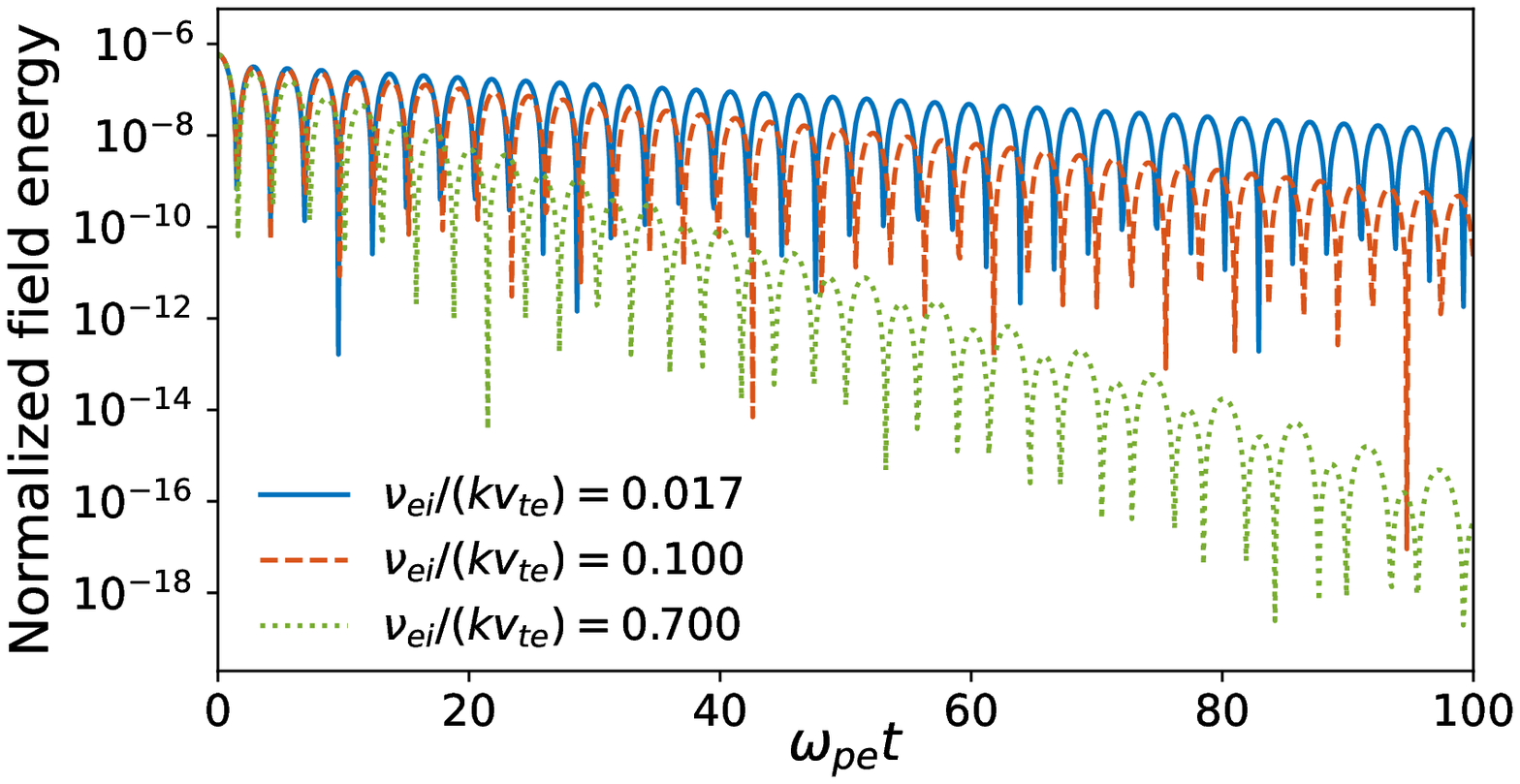}
  \end{subfigure}
  \begin{subfigure}[b]{0.495\textwidth}
    \includegraphics[width=\textwidth]{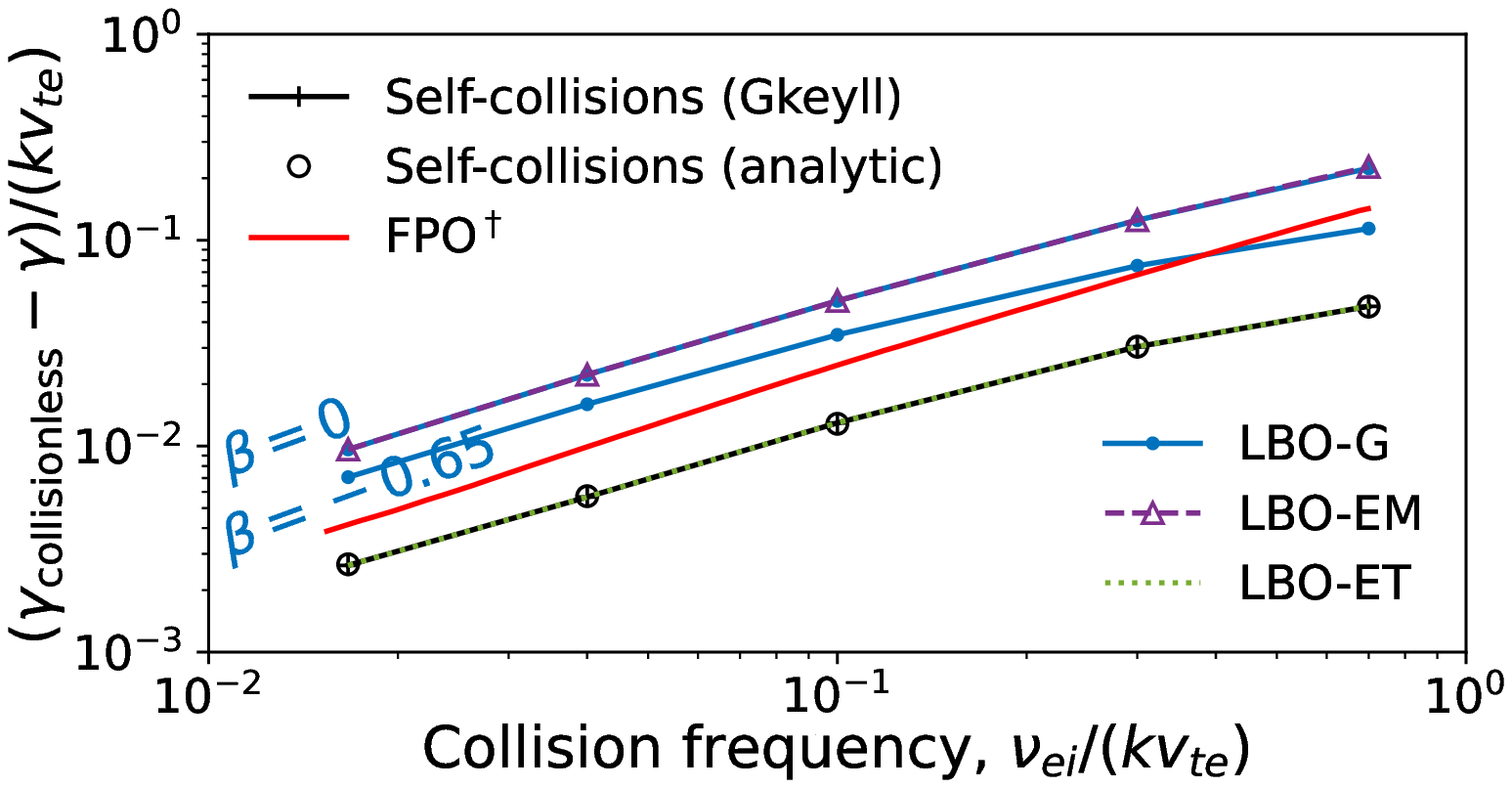}
  \end{subfigure}
  \caption{Left: normalized volume integrated field energy in simulations of electron Langmuir waves as they Landau damp over time. Right: damping rates of electron Langmuir waves given as the offset from the collisionless value of $\gamma_{\mathrm{collisionless}}/\omega_{pe}=-1.247\times10^{-2}$ for the LBO-G with $\beta=0$ and $\beta=-0.65$ (blue circles), the LBO-EM (purple triangles) and the LBO-ET (dotted green) compared to those with a full FPO~\citep{Jorge2019a} (solid red). If using only self-collisions, numerical \gkeyll~data (black crosses) agrees with theory (black circles).}
  \label{fig:landau}
\end{figure}

As the simulation proceeds we see the amplitude of the electrostatic wave damp, which can be appreciated by examining the volume integrated field energy over time as shown in figure~\ref{fig:landau}(left). We can quantify the rate at which these waves damp and plot it as a function of collision frequency as is done in figure~\ref{fig:landau}(right). If one were to only use self-species collisions one would obtain the results shown with black crosses, and for that case the equations are sufficiently simple that one can obtain an analytic dispersion relation~\citep{Anderson2007,Francisquez2020} which agrees well with the numerical results (black circles), providing additional confidence in the \gkeyll~implementation. When we introduce electron-ion collisions obtaining analytic growth rates is more difficult. So we instead compare the results obtained with the LBO-G (solid blue), the LBO-EM (dashed purple) and the LBO-ET (dotted green) with previously reported results for the FPO~\citep{Jorge2019a}.

The LBO-G simulations were performed using $\nuei^G=\sqrt{2}\nu_{ee}$ ($=(m_i/m_e)\nuie$) since this is the relationship assumed in the reference FPO work~\citep{Jorge2019a}. Figure~\ref{fig:landau}(right) suggests that the LBO-G can provide a more accurate description of this kinetic phenomenon than, say, using self-species collisions only. There is the caveat however that we have not established from first principles what the most suitable choice of the free parameter $\beta$ ought to be at any given collision frequency. We scanned this parameter and show the results for $\beta=0$ and $\beta=-0.65$, the latter bringing the damping rates closer to those of the full FPO. But it is apparent that the wrong choice of $\beta$ can also result in significant deviation from the FPO.

Also shown in figure~\ref{fig:landau}(right) are the damping rates obtained when using the LBO-EM and the LBO-ET. Despite having a different model for $\usr$ and $\vtsr$, LBO-EM has the same collision frequency ($\nuei^M=\nuei^G(\beta=0)$) and gives the same damping rates as the LBO-G with $\beta=0$ (top blue and dashed purple lines). The LBO-ET on the other hand has a collision frequency that is smaller by the mass ratio ($\nuei^T = m_e\nuei^M/(m_e+m_i)\simeq (m_e/m_i)\nuei^M$), and therefore is essentially equivalent to neglecting cross-species collisions for this problem; i.e. solid black and dotted green lines agree. If we were to run the simulation with the LBO-ET but the same value of $\nuei$ as the LBO-EM then we would simply obtain the same results as if we had used the LBO-EM.

\subsection{Velocity and temperature relaxation} \label{sec:relaxTest}

As a final benchmark of the multispecies LBO algorithms and solvers we employ the gyroaveraged LBO to model the relaxation of a deuterium plasma to thermal equilibrium. We employ identical conditions, as best as we can tell, to those used in a benchmark of the FPO in the XGC code~\citep{Hager2016}. The same test was recently performed with a finite-volume implementation of the LBO-ET and LBO-EM in GENE-X~\cite{Ulbl2021}. This means that the initial distribution functions are described by the bi-Maxwellians:
\begin{equation}
    \fs\left(\vpar,\mu,t=0\right) = \frac{n_0}{\alpha\left(2\pi\vts^2\right)^{3/2}}\exp\left[-\frac{\left(\vpar-\upars\right)^2+2\mu B/(\ms\alpha)}{2\vts^2}\right]
\end{equation}
where $B=1$ T, $\vts=\sqrt{T_{\sind 0}/\ms}$ $\alpha=1.3$, $\upari=50(m_e/m_i)\vti$, $\upare=0.5\sqrt{m_e/m_i}\vts$, $T_{i0}=200$ eV and $T_{e0}=300$ eV. Note that these reference temperatures are slightly different than the true initial temperatures $T_i(t=0)=240$ eV and $T_e(t=0)=360$ eV given by $T_s=(2T_{\perp s} + T_{\parallel s})/3$. For this test we once again neglect the collisionless terms and use a collision frequency that depends on time, i.e. $\nusr=\nusr(T_s(t),T_r(t))$. The phase space $\left[-2,2\right]\times\left[-5\vts,5\vts\right]\times\left[0,\ms(5\vts)^2/(2B)\right]$ is meshed with $1\times16^2$ cells and dynamic fields are expanded in a piecewise linear ($p=1$) basis. This resolution and velocity-space extents were confirmed as sufficient by convergence tests.

\begin{figure}
  \centering
  \begin{subfigure}[b]{0.49\textwidth}
    \centering
    \includegraphics[width=\textwidth]{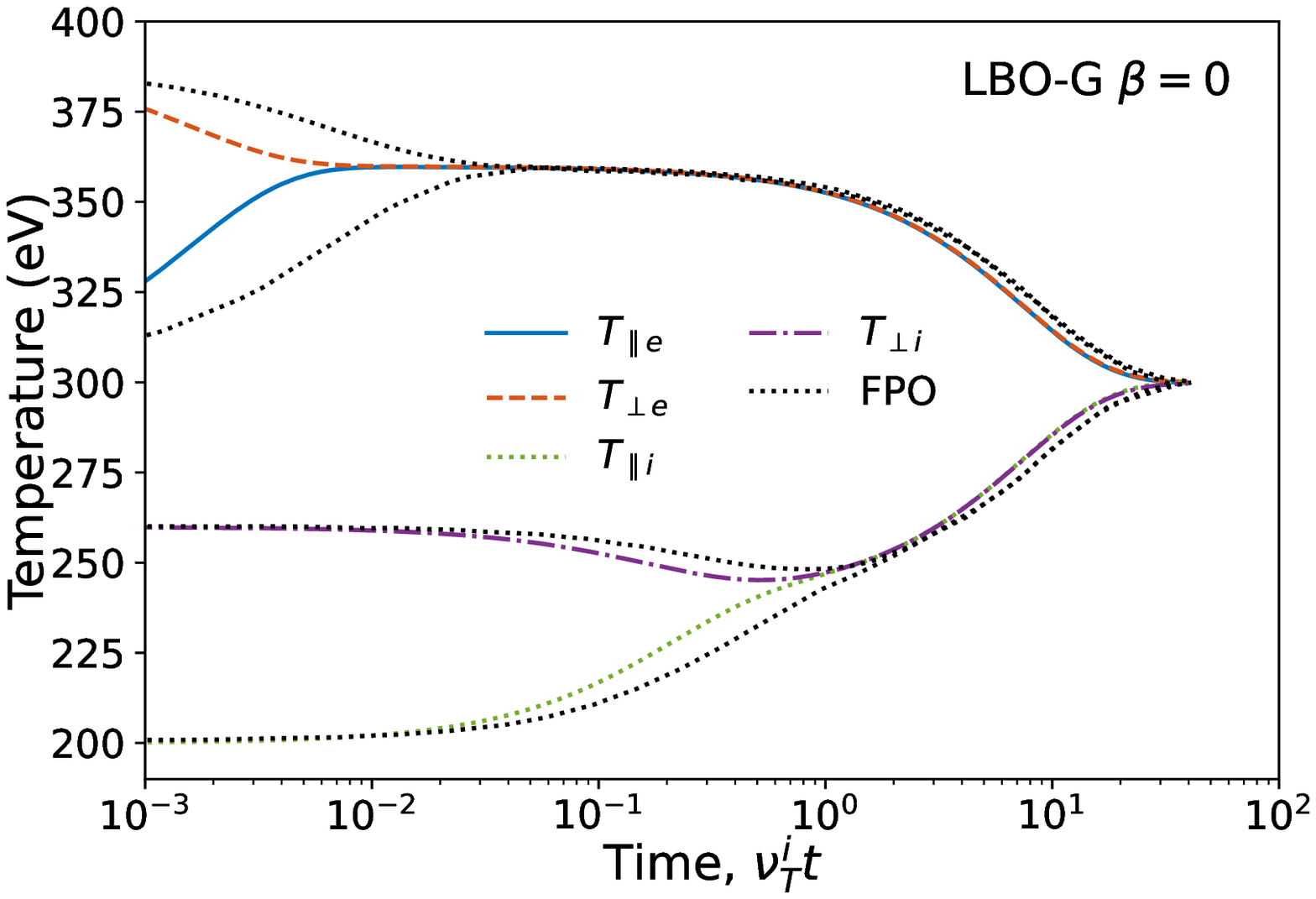}
  \end{subfigure}
  \begin{subfigure}[b]{0.49\textwidth}
    \centering
    \includegraphics[width=\textwidth]{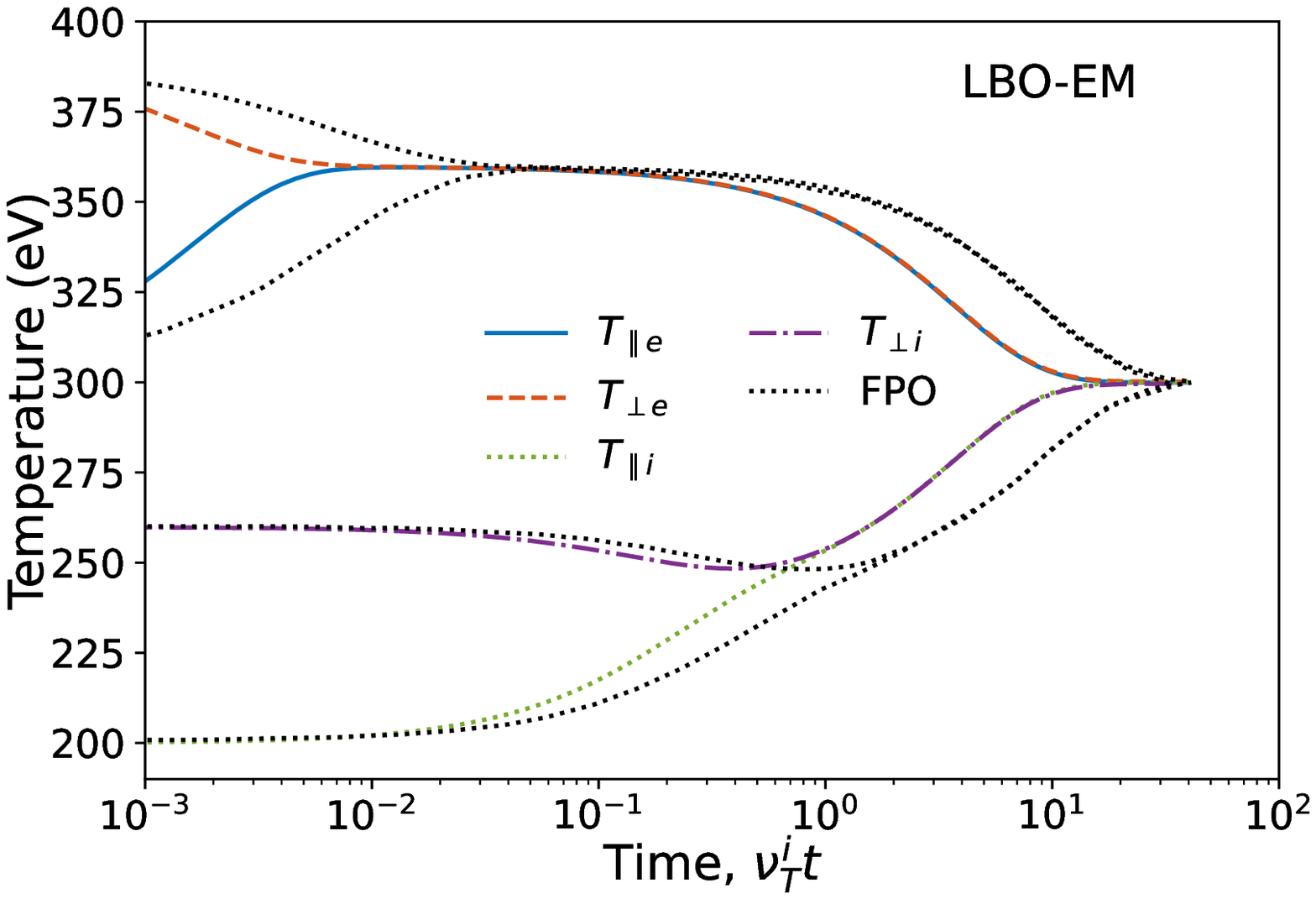}
  \end{subfigure}
  \begin{subfigure}[b]{0.49\textwidth}
    \centering
    \includegraphics[width=\textwidth]{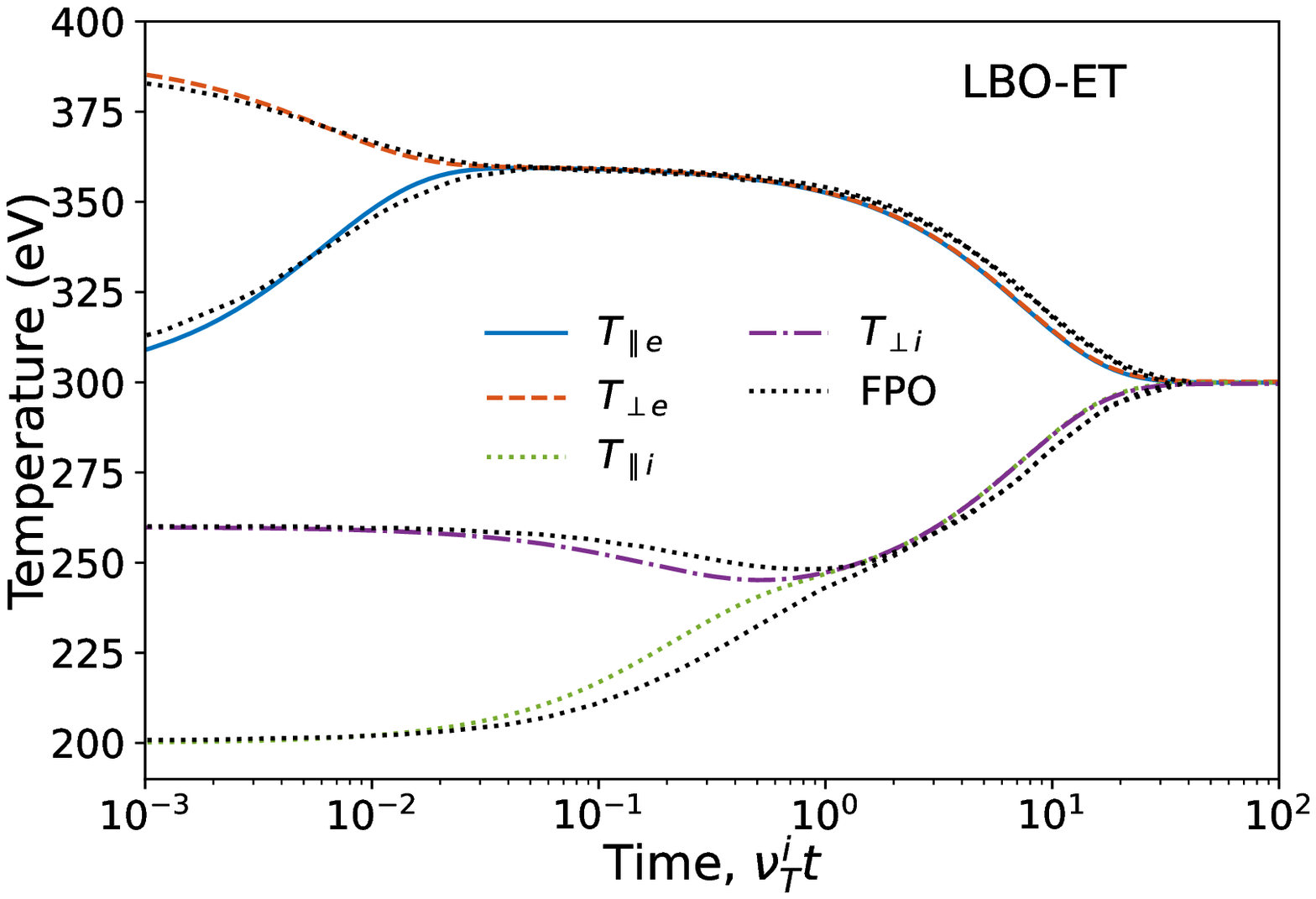}
  \end{subfigure}
  \caption{Isotropization and relaxation of temperatures as deuterium ions and electrons collide with themselves and each other, compared to the results from the full FPO~\citep{Hager2016}. Top left: LBO-G with $\beta=0$. Top right: LBO-EM. Bottom: LBO-ET.}
  \label{fig:hagerTempRelax}
\end{figure}

\begin{figure}
  \centering
  \begin{subfigure}[b]{0.49\textwidth}
    \centering
    \includegraphics[width=\textwidth]{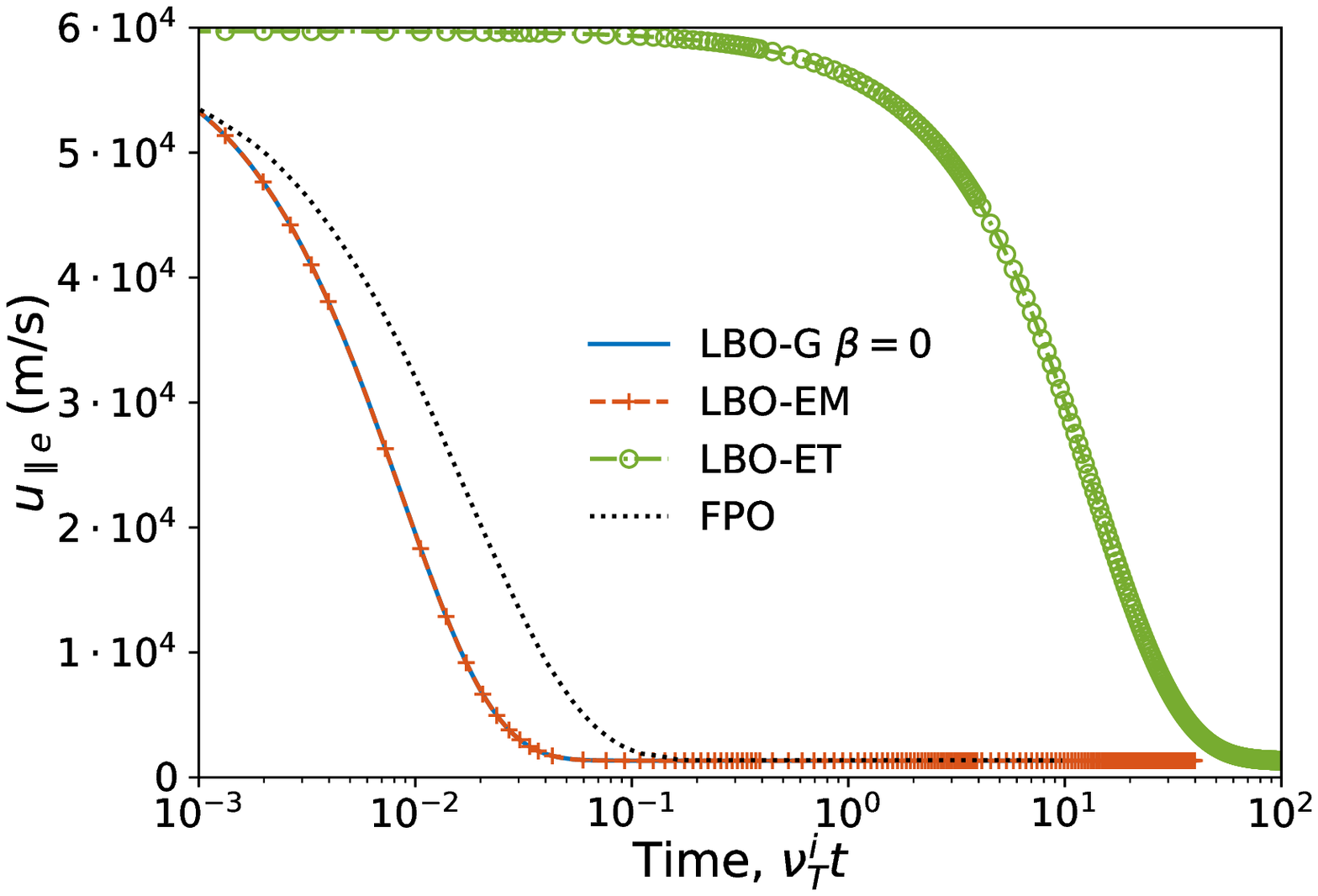}
  \end{subfigure}
  \begin{subfigure}[b]{0.49\textwidth}
    \centering
    \includegraphics[width=\textwidth]{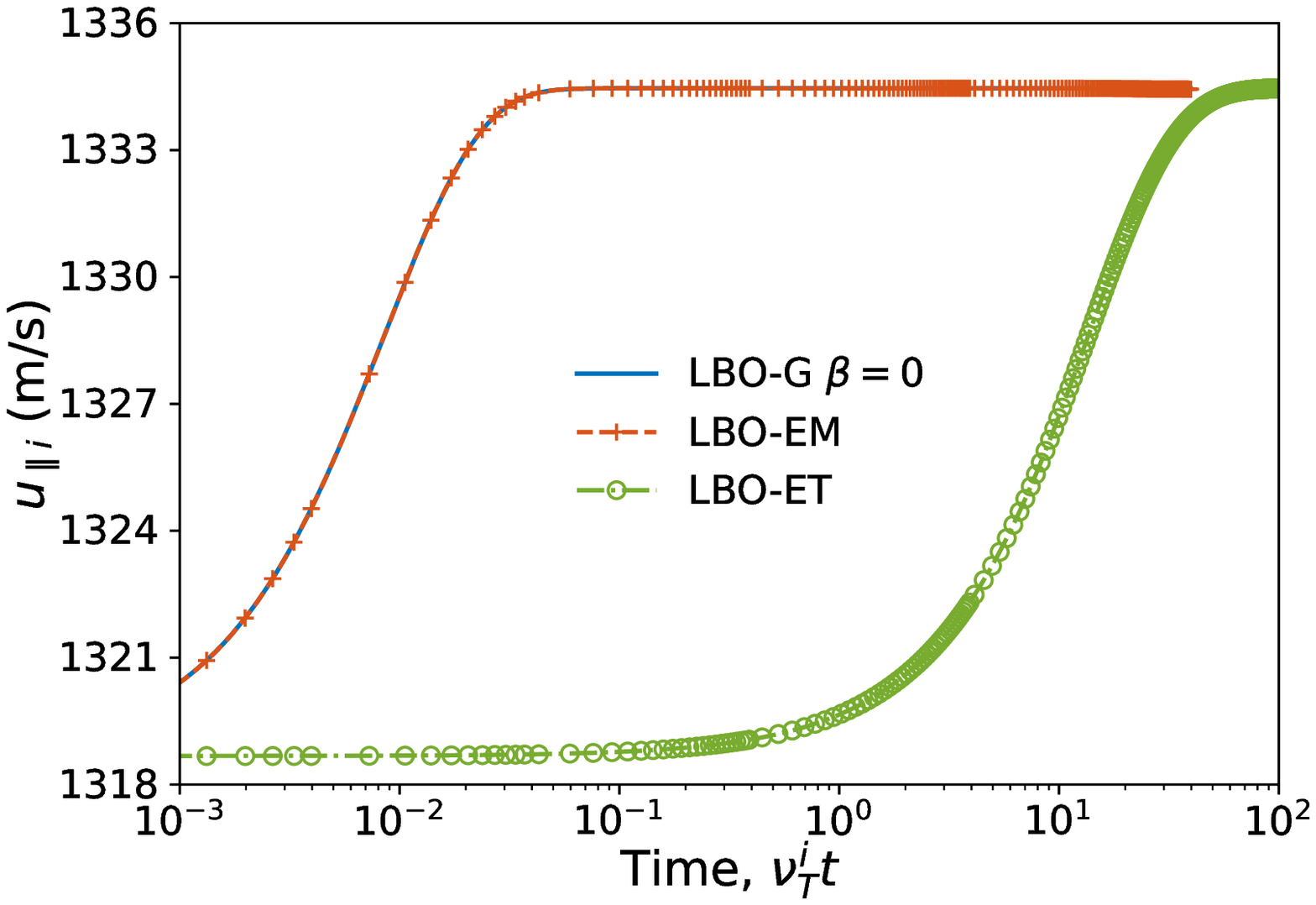}
  \end{subfigure}
  \caption{Relaxation of parallel flow speed as deuterium ions and electrons collide with themselves and each other. Electron flow speeds are compared to the results from the full FPO (dotted black,~\cite{Hager2016}). Flow speeds when using the LBO-G ($\beta=0$) are given in solid dark blue, results for the LBO-EM in dashed orange with crosses, and green dash-dot lines with circles represent results obtained with the LBO-ET.}
  \label{fig:hagerUparRelax}
\end{figure}

Figure~\ref{fig:hagerTempRelax} provides the time evolution of the parallel and perpendicular temperatures for the LBO-G ($\beta=0$), LBO-EM and LBO-ET operators compared to the previously reported FPO results\footnote{Note that the FPO results here (and those in figure~\ref{fig:hagerUparRelax}) have been shifted in time by $-\Delta t=3.85783\times10^{-7}$ s compared to those in~\cite{Hager2016}, since that work shifted them by $\Delta t$ in order to show the results with a logarithmic $x$-axis.}~\citep{Hager2016}. The first event is the isotropization of the electrons followed by the isotropization of the ions, happening on the $\nu_{ss}^{-1}$ time scale. We used
\begin{equation}
    \nu_{ss} = \frac{1}{\sqrt{2}}\frac{q_{\sind}^4\ns\log\Lambda_{ss}}{3\left(2\pi\right)^{3/2}\epsilon_0^2\ms^2\vts^3},
\end{equation}
for the like-species scattering rate. Note that the LBO-G and LBO-EM exhibit a delayed isotropization time compared to the FPO's, an observation that~\cite{Pezzi2015} had also made while comparing the self-species Dougherty operator to the FPO (in Landau form). Later, on the $\nuie^{-1}$ time scale we see the electrons and ions come into thermal equilibrium with each other, a process that is better described by both the LBO-G and the LBO-ET operators since, after all, the LBO-EM made no attempt at matching the FPO thermal relaxation rates. The time-axis on these plots has been normalized to the isotropization rate~\citep{Huba2013}
\begin{equation}
    \nu_T^i = \frac{2\sqrt{\pi}q_i^4n_i\log\Lambda_{ii}}{\left(4\pi\epsilon_0\right)^2m_i^2\vti^3}A^{-2}\left[-3+\left(A+3\right)a_F\right],
\end{equation}
where $\vti$ and $\log\Lambda_{ii}$ use the initial ion temperature (240 eV), and $A=T_{\perp i}/T_{\parallel i}-1$ and $a_F=A^{-1/2}\tan^{-1}(A^{1/2})$ if $A>0$ or $a_F=(-A)^{1/2}\tanh^{-1}(-A)^{1/2}$ if $A<0$.

We can also examine the velocity evolution as the plasma approaches an equilibrium as is done in figure~\ref{fig:hagerUparRelax}. The first thing we notice is that the LBO-ET (green dash-dot with circles) grossly overestimates the time-scale on which the electron flow relaxes to the ion flow, which happens because the ions are so much more massive. By definition the LBO-ET did not attempt to match the momentum relaxation rate, and the we see the result of that here. On the other hand, the LBO-G (solid blue) and LBO-EM (dashed orange with crosses) models do a better job of approximating the FPO results for the slowing down of electrons, since their formulation included matching the FPO's momentum relaxation rates. There's still a discrepancy, e.g. between solid blue and dashed orange lines, although we point out that the LBO-G and LBO-EM would appear to match the analytic result based on the flow relaxation frequency given by the friction force at large mass ratio~\citep{Hinton1976} (see figure 4 of~\cite{Hager2016}).

\section{Conclusion} \label{sec:conclusion}

This work presented three separate formulations of full-$f$ nonlinear multispecies collisions based on the model Lenard-Bernstein or Dougherty operator (LBO), following the ideas~\cite{Greene1973} and~\cite{Haack2017} employed for the BGK operator. This resulted in the LBO-G, LBO-EM and LBO-ET operators, each providing different formulas for the cross-species primitive moments $\usri$ and $\vtsr$ and collision frequency $\nusr$. The LBO-G attempts to exactly match the thermal and momentum relaxation rates of the Fokker-Planck operator (FPO), but it introduces a free parameter $\beta$. The LBO-EM only matches the FPO momentum relaxation rate, while the LBO-ET only tries to approximately match the FPO thermal relaxation rate. Gyroaveraged versions of this operators were also provided in this work, which may be used in long-wavelength gyrokinetic models. Compared to previous works, the multispecies LBO model presented here has the following advantages:
\begin{itemize}
    \item It is suitable for arbitrary mass ratios.
    \item Some pathologies, such as negative cross-species temperatures (possible in the BGK operator of~\cite{Greene1973}), are avoided.
    \item It conserves energy and momentum exactly.
    \item It approximately reproduces the FPO's momentum and thermal relaxation rates.
    \item A proof of non-decreasing entropy (the \boltzH~theorem) exists.
\end{itemize}

These multispecies LBO models may also be discretized for numerical implementation using a discontinuous Galerkin (DG) method in the spirit of~\cite{Hakim2020} and~\cite{Francisquez2020}. We provided an algorithm for a DG discretization of such operators based on weak projections and the recovery of discontinuous derivatives across cell boundaries~\citep{Hakim2020}. The primary focus of this work was, however, the computation of the cross-primitive moments $\usri$ and $\vtsr$ in a manner that results in an exactly conservative algorithm, i.e. capable of conserving particle, momentum and kinetic energy density independently of resolution. This property was accomplished by solving a weak system of equations consisting of the discrete equivalent of momentum and energy conservation, and in the case of the LBO-G, a discrete equivalent of the momentum and thermal relaxation rate constraints. Discrete conservation was also attained when piecewise-linear basis functions ($p=1$) were used by carefully employing the projection of $v^2$ onto the basis (or $\vpar^2$ for the gyroaveraged operator).

Our tests indicate that the implementation in \gkeyll~exhibits this exact conservation feature, for all the velocity dimensions and polynomial orders tested. Exact conservation was also confirmed in \gkeyll's gyroaveraged solver for one and two velocity dimensions. In addition we combined the LBO solver with \gkeyll's Vlasov-Maxwell solver and examined the impact that LBO cross-species collisions has on the Landau damping rates of electron Langmuir waves. Due to the definition of the LBO-ET collision frequency, such operator gave no improvements over using self-species collisions only, while the LBO-G and LBO-EM gave slightly more accurate descriptions of this phenomenon. The LBO-G can be made to agree more with the FPO by choosing a different value of $\beta$, but we have not presented a first-principles model for that free parameter yet. Despite this unspecified parameter, the LBO-G operator has been in use by \gkeyll's Vlasov and gyrokinetic solvers for quite some time now. For example, recent Vlasov-Maxwell simulations using this operator showed the inhibition of magnetic dynamo due to Landau damping~\citep{Pusztai2020}. Nevertheless, this $\beta$ parameter will be the focus of follow up work.

Lastly, we benchmarked the gyroaveraged multispecies LBO by simulating a system in which ions and electrons are anisotropic, drifting relative to each other, and out of thermal equilibrium. The LBO-EM and LBO-ET each do better at approximating the FPO's velocity and temperature evolution, as their formulation would predict. The LBO-G is perhaps the best choice here, since it does well at matching the temperature evolution and provides the same level of accuracy when it comes to the velocity relaxation as the LBO-EM.

\section*{Acknowledgements}

We express our gratitude towards Rogerio Jorge and Robert Hager for clarifying how the FPO results were obtained, as well as our gratitude for the other members of the \gkeyll~team who aided this work. We used the Stellar cluster at Princeton University and the Cori cluster at the National Energy Research Scientific Computing Center (NERSC), a U.S. Department of Energy Office of Science User Facility. M.F., A.H. and G.W.H. were supported by the Partnership for Multiscale Gyrokinetic Turbulence (MGK) and the High-Fidelity Boundary Plasma Simulation (HBPS) projects, part of the U.S. Department of Energy (DOE) Scientific Discovery Through Advanced Computing (SciDAC) program, and the DOE's ARPA-E BETHE program, via DOE contract DE-AC02-09CH11466 for the Princeton Plasma Physics Laboratory. J.J was supported by a NSF Atmospheric and Geospace Science Postdoctoral Fellowship (Grant No. AGS-2019828). M.F., as well as D.R.E., was also supported by the Partnership for Multiscale Gyrokinetic Turbulence (MGK) (subaward No. UTA18-000276 to M.I.T. under U.S. DOE Contract DE-SC0018429).

\section*{Declaration of Interests}
The authors report no conflict of interest.

\section*{Data availability statement}
The data that support the findings of this study are openly available in Zenodo at \url{https://doi.org/10.5281/zenodo.6350748}.

\appendix

\section{Getting \gkeyll~and reproducing results} \label{sec:getGkeyll}

Readers may reproduce our results and also use Gkeyll for their applications. The code and input files used here are available online. Full installation instructions for Gkeyll are provided on the \gkeyll~website~\cite{gkeyllWeb}. The code can be installed on Unix-like operating systems (including Mac OS and Windows using the Windows Subsystem for Linux) either by installing the pre-built binaries using the conda package manager (\url{https://www.anaconda.com}) or building the code via sources. The input files used here are under version control and can be found at \url{https://github.com/ammarhakim/gkyl-paper-inp/tree/master/2021_JPP_crossLBO}.

\section{\boltzH-theorem proof} \label{sec:entrApp}

In this section we show that the improved interspecies Dougherty collisions do not decrease total entropy given the cross-species primitive moments for collisions between species $\sind$ and species $\rind$ (equations~\ref{eq:usrBetaSym}-\ref{eq:TsrBetaSym}). The rate of change of the entropy $\entr$ can be written as
\begin{eqnal} \label{eq:Hdot1}
\pd{\entr}{t} &= -\pd{}{t}\sum_\sind\int\text{d}^\vdim v~\fs\ln \fs \\
&= -\sum_\sind\int\text{d}^\vdim v~\pd{\fs}{t}\left(\ln \fs+1\right), \\
&= -\sum_\sind\int\text{d}^\vdim v~\nusr\left(\divv{\vJsr}\right)\left(\ln \fs+1\right),
\end{eqnal}
where $\vJsr = \left(\vv-\vusr\right)\fs+\vtsr^2\gradv{\fs}$. Integrate equation~\ref{eq:Hdot1} by parts and use the fact that $\fs\to0$ faster than any polynomial or logarithmic singularity. In the interest of simplicity we adopt the notation $\intvdim=\int\text{d}^\vdim v$, then the time rate of change $\dot{\entr}$ becomes
\begin{eqnal}
\dot{\entr} &= \sum_\sind\intvdim\nusr\vJsr\,\gradv{\left(\ln \fs+1\right)} \\
&= \sum_\sind\intvdim\nusr\left[\left(\vv-\vusr\right)\fs+\vtsr^2\gradv{\fs}\right]\cdot\gradv{\ln\fs}, \\
&= \sum_\sind\intvdim\nusr\left[\left(\vv-\vusr\right)\cdot\gradv{\fs} +\vtsr^2\gradv{\fs}\cdot\gradv{\ln\fs}\right].
\end{eqnal}
The first term can be integrated again so that upon discarding the surface term, and adopting the notation $\dotNsr=\nusr\vdim\ns$ and $\dotTsr=\nusr\vtsr^2$, one obtains
\begin{eqnal} \label{eq:easyHdot}
\dot{\entr} &= \sum_\sind\left(-\dotNsr+\dotTsr\intvdim\gradv{\fs}\cdot\gradv{\ln\fs}\right).
\end{eqnal}

At this point we can ask what is the distribution function that minimizes $\dot{\entr}$. Given a set of primitive moments ($\usri$, $\vtsr$, and also $\usi$, $\vts$) and the virtual displacement $\delta \fs = \fs-\fsZ$, the response of the functional in equation~\ref{eq:easyHdot} is
\begin{eqnal}
\delta\dot{\entr} &= \sum_\sind\dotTsr\intvdim\left[\delta\left(\frac{1}{\fs}\right)\left|\gradv{\fsZ}\right|^2+\frac{1}{\fsZ}\delta\left|\gradv{\fs}\right|^2\right], \\
&\approx \sum_\sind\dotTsr\intvdim\frac{1}{\fsZ}\left(-\frac{\delta \fs}{\fsZ}\left|\gradv{\fsZ}\right|^2+2\gradv{\fsZ}\cdot\gradv{\delta \fs}\right), \\
&= \sum_\sind\dotTsr\intvdim\left\lbrace\frac{1}{\fsZ}\left(-\frac{\delta \fs}{\fsZ}\right)\left|\gradv{\fsZ}\right|^2 +2\left[\divv{\frac{\delta \fs}{\fsZ}\gradv{\fsZ}}-\delta \fs\divv{\frac{1}{\fsZ}\gradv{\fsZ}}\right]\right\rbrace.
\end{eqnal}
The second term vanishes since $\delta \fs\to\pm\infty$ as $v_i\to\pm\infty$. Thus at an extremum
\begin{eqnal} \label{eq:HdotNew}
\delta\dot{\entr} &= \sum_\sind\dotTsr\intvdim\left[-\frac{1}{\fsZ^2}\left|\gradv{\fsZ}\right|^2-2\divv{\frac{1}{\fsZ}\gradv{\fsZ}}\right]\delta\fs
\end{eqnal}
must vanish, and since equation~\ref{eq:easyHdot} has no upper bound this extremum must be a minimum. At this point we can impose the conditions
\begin{align} 
&\intvdim\delta \fs =0, \label{eq:restrict1}\\
&\intvdim\ms \vv\thinspace\delta \fs =0, \label{eq:restrict2}\\
&\intvdim\frac{1}{2}\ms v^2\thinspace\delta \fs =0,\label{eq:restrict3}
\end{align}
requiring the virtual displacement to not alter the moments of the solution (but does not mean that the moments are constant in time). From equations~\ref{eq:HdotNew}-\ref{eq:restrict3} we can deduce that for equation~\ref{eq:HdotNew} to vanish for all displacements $\delta \fs$ it must be that
\begin{eqnal} \label{eq:interH}
\left|\gradv{\ln \fsZ}\right|^2+2\lapv{\ln \fsZ} = a + \vb\cdot\vv+c v^2,
\end{eqnal}
where $a$, $\vb$ and $c$ are constants. We can re-write this equation as
\begin{eqnal} \label{eq:newHeq}
h(t,\vx,\vv)^2+2\divv{\vh(t,\vx,\vv)} = a + \vb\cdot\vv+c v^2,
\end{eqnal}
where $\vh(t,\vx,\vv)=\gradv{\ln \fsZ}$. We claim that the solution to this nonlinear inhomogeneous equation is
\begin{equation} \label{eq:solution}
\vh(t,\vx,\vv) = \vhO + h_1\vv,
\end{equation}
with $\vhO$ and $h_1$ yet undetermined constants. Check by substituting equation~\ref{eq:solution} into equation~\ref{eq:newHeq}:
\begin{eqnal}
h(t,\vx,\vv)^2 + 2\divv{\vh(t,\vx,\vv)} &= h_0^2 + 2h_1\vhO\cdot\vv + h_1^2v^2 + 2\divv{\left(\vhO + h_1\vv\right)},
\end{eqnal}
which has the same form as the right side of equation~\ref{eq:newHeq} with $a=h_0^2+2\vdim h_1$, $\vb=2h_1\vhO$ and $c=h_1^2$. Going back to the definition of $\vh(t,x_i,\vi)$, we can arrive at
\begin{eqnal} \label{eq:minF}
\ln \fsZ &= g_0 + \vhO\cdot\vv+\frac{1}{2}h_1v^2, \\
\Rightarrow \fsZ &= A \exp\left(\vhO\cdot\vv+\frac{1}{2}h_1v^2\right).
\end{eqnal}

We can now explore whether our minimized $\dot{\entr}$ falls below zero. For this we rewrite equation~\ref{eq:easyHdot} making use of vanishing total derivatives
\begin{equation}
\min\left(\dot{\entr}\right) = \sum_\sind\left(-\dotNsr-\dotTsr\intvdim\fsZ\divv{\frac{1}{\fsZ}\gradv{\fsZ}}\right),
\end{equation}
and, from equation~\ref{eq:interH},
\begin{eqnal}
\fsZ\divv{\frac{1}{\fsZ}\gradv{\fsZ}} = \frac{a + \vb\cdot\vv+c v^2}{2}\fsZ-\frac{1}{2\fsZ}\left|\gradv{\fsZ}\right|^2.
\end{eqnal}
Putting these two equations together we have:
\begin{eqnal} \label{eq:cMinH}
&\min\left(\dot{\entr}\right) = \sum_\sind\left[-\dotNsr-\frac{1}{2}\dotTsr\intvdim\left(a + \vb\cdot\vv+c v^2-\left|\gradv{\ln \fsZ}\right|^2\right)\fsZ\right].
\end{eqnal}
Take the derivative of equation~\ref{eq:minF} and insert it into equation~\ref{eq:cMinH} to obtain
\begin{eqnal}
&\min\left(\dot{\entr}\right) = \sum_\sind\left[-\dotNsr-\frac{1}{2}\dotTsr\intvdim\left(a + \vb\cdot\vv+c v^2-\left|\v{h_0}+h_1\vv\right|^2\right)\fsZ\right].
\end{eqnal}
Employing the definitions of $a$, $\bi$ and $c$ above this becomes
\begin{eqnal} \label{eq:Hlatest}
\min\left(\dot{\entr}\right) &= \sum_\sind\left(-\dotNsr-\vdim h_1\dotTsr\intvdim\fsZ\right).
\end{eqnal}
If we requite that the zeroth moment of $\fsZ$ equals $\ns$, we find that
\begin{eqnal}
\fsZ &= \ns\left(-\frac{h_1}{2\pi}\right)^{\vdim/2}\exp\left[\frac{\left(\v{h_0}+h_1\vv\right)^2}{2h_1}\right].
\end{eqnal}
Identify $h_1$ with $-\vts^{-2}$ such that the minimizing function becomes
\begin{eqnal}
\fsZ = \frac{\ns}{\left(2\pi\vts^2\right)^{\vdim/2}}\exp\left[-\frac{\left(\vts^2\v{h_0}-\vv\right)^2}{2\vts^2}\right],
\end{eqnal}
and by taking the first moment of this distribution it would become clear that $\vts^2\vhO = \vus$. The minimizing distribution is thus a Maxwellian with number density $\ns$, mean flow velocity $\vus$ and thermal speed $\vts$. Since we have found a single distribution that minimizes $\dot{\entr}$ then the minimum given below must be global.

One can show that if the two colliding distributions are Maxwellian, that the total entropy does not decrease. We can check this here by going back to equation~\ref{eq:Hlatest}, and find that the minimum entropy rate of change is
\begin{eqnal}
\min\left(\dot{\entr}\right) &= -\vdim \sum_\sind \ns\nusr\left(1 - \frac{\vtsr^2}{\vts^2}\right).
\end{eqnal}
At this point one must substitute the definition for the cross-species thermal speed in equation~\ref{eq:TsrBetaSym} to yield
\begin{eqnal}
&\min\left(\dot{\entr}\right) = \vdim \sum_\sind \frac{\ns\nusr}{\vts^2}\frac{\fnus}{2}\frac{1+\beta}{1+\frac{\mr}{\ms}}\left[\frac{\mr}{\ms}\vtr^2+\frac{1}{\vdim}\frac{\mr}{\ms}\left(\vus-\vur\right)^2-\vts^2\right], \\
&\quad= \vdim \frac{\ms\ns\nusr}{\vts^2}\frac{\fnus}{2}\frac{1+\beta}{\ms+\mr}\sum_\sind \left[\frac{\Tr}{\Ts}+\frac{1}{\vdim}\frac{\mr}{\ms}\frac{\left(\vus-\vur\right)^2}{\vts^2}-1\right], \\
&\quad= \vdim \frac{\ms\ns\nusr}{\vts^2}\frac{\fnus}{2}\frac{1+\beta}{\ms+\mr}\left[\frac{\left(\Tr-\Ts\right)^2}{\Ts\Tr}+\frac{1}{\vdim}\left(\frac{\mr}{\ms}\frac{1}{\vts^2}+\frac{\ms}{\mr}\frac{1}{\vtr^2}\right)\left(\vus-\vur\right)^2\right] \geq 0.
\end{eqnal}
and thus the entropy cannot decrease and the \boltzH-theorem of this nonlinear full-$f$ multi-species collision model is guaranteed.

\section{Energy conservation with piecewise linear basis} \label{sec:pOenerConserv}

\subsection{Cartesian p=1 energy conservation} \label{sec:pOenerConservVM}

The derivation of a constraint on the operator to conserve energy in section~\ref{sec:discreteEnerConserv} relied on $v^2$ belonging to the space span by the basis set. For piecewise linear basis ($p=1$) that is not the case, so instead we can guarantee that the algorithm preserves the projection of the energy onto the basis. We use the notation and strategy first outlined in~\cite{Hakim2020} for self-species collisions; $\vSqBar$ is the projection of $v^2$ onto the basis. For the energy to be conserved the left side of equation~\ref{eq:dgScheme} has to be zero after making the substitution $\psi_\ell=\ms\vSqBar/2$, summing over species and over all cells. Those steps lead to the relation
\begin{eqnal} \label{eq:pOEner}
&\sum_j\int_{x_{i-1/2}}^{x_{i+1/2}}\left\{\frac{\ms}{2}\nusr\left[\left(\vSqBar \Gs-\pd{\vSqBar}{v}\vtsr^2\hat{\fs}\right)\Bigg|_{v_{j-1/2}}^{v_{j+1/2}} - \int_{v_{j-1/2}}^{v_{j+1/2}} \pd{\vSqBar}{v}\left(v-\usr\right)\fs\dv\right] \right.\\
&\left.\quad+ \frac{\ms}{2}\nurs\left[\left(\vSqBar \Gr-\pd{\vSqBar}{v}\vtrs^2\hat{\fr}\right)\Bigg|_{v_{j-1/2}}^{v_{j+1/2}} - \int_{v_{j-1/2}}^{v_{j+1/2}} \pd{\vSqBar}{v}\left(v-\urs\right)\fr\dv\right]\right\}\dx = 0,
\end{eqnal}
where we used $\partial^2\vSqBar/\partial v^2=0$. Next we use the fact that~\citep{Hakim2020}
\begin{equation} \label{eq:vSqBar_x}
    \frac{1}{2}\pd{\vSqBar}{v} = \vcen_j = \frac{v_{j-1/2}+v_{j+1/2}}{2}
\end{equation}
(the cell center) and the continuity of $\vSqBar\Gs$ as well at the zero flux boundary conditions, to turn equation~\ref{eq:pOEner} into
\begin{eqnal}
&\int_{x_{i-1/2}}^{x_{i+1/2}}\sum_j\left\{\ms\nusr\left[-\vcen_j\vtsr^2\hat{\fs}\Big|_{v_{j-1/2}}^{v_{j+1/2}} - \int_{v_{j-1/2}}^{v_{j+1/2}} \vcen_j\left(v-\usr\right)\fs\dv\right] \right.\\
&\left.\quad+ \ms\nurs\left[-\vcen_j\vtrs^2\hat{\fr}\Big|_{v_{j-1/2}}^{v_{j+1/2}} - \int_{v_{j-1/2}}^{v_{j+1/2}} \vcen_j\left(v-\urs\right)\fr\dv\right]\right\}\dx = 0.
\end{eqnal}
Carrying out the velocity integrals this equation becomes
\begin{eqnal}
&\int_{x_{i-1/2}}^{x_{i+1/2}}\left\{\ms\nusr\left[-\vtsr^2\left(\vcen_j\fs(v_{j\pm1/2})\Big|_{\jmin}^{\jmax}-\MZstars\right) - \left(\MTstars-\usr\MOstars\right)\right] \right.\\
&\left.\quad+ \ms\nurs\left[-\vtrs^2\left(\vcen_j\fr(v_{j\pm1/2})\Big|_{\jmin}^{\jmax}-\MZstarr\right) -  \left(\MTstarr-\urs\MOstarr\right)\right]\right\}\dx = 0,
\end{eqnal}
where the $\pm$ sign is used when evaluating at $\jmax/\jmin$ and we have introduced the star moments
\begin{eqnal}
    \MZstars &= \sum_{j=1}^{\Nv-1}\left(\vcen_{j+1}-\vcen_j\right)\hat{f}_{\sind,j+1/2}, \\
    \MOstars &= \sum_{j=1}^{\Nv}\int_{v_{j-1/2}}^{v_{j+1/2}}\vcen_j\fs\,\dv, \\
    \MTstars &= \sum_{j=1}^{\Nv}\int_{v_{j-1/2}}^{v_{j+1/2}}\vcen_jv\fs\,\dv.
\end{eqnal}
Therefore our DG scheme will conserve energy if we enforce the following weak constraint
\begin{eqnal}
&\ms\nusr\left[\usr\MOstars + \vtsr^2\left(\MZstars-\vcen_j\fs(v_{j\pm1/2})\Big|_{\jmin}^{\jmax}\right)\right] \\
&\quad+ \mr\nurs\left[\urs\MOstarr + \vtrs^2\left(\MZstarr-\vcen_j\fr(v_{j\pm1/2})\Big|_{\jmin}^{\jmax}\right)\right] \doteq \ms\nusr\MTstars + \mr\nurs\MTstarr.
\end{eqnal}

In order to formulate an energy-conserving LBO-G with $p=1$ we also need to re-examine the thermal relaxation rate of the discrete operator, equation~\ref{eq:discThermalRelaxRate}. If we instead substitute $\psi_\ell=\ms\overline{\left(v-\us\right)^2}/2=\ms\left(\vSqBar-2v\us+\us^2\right)/2$ into equation~\ref{eq:dgScheme} and sum over velocity-space cells we get
\begin{eqnal}
\sum_j\int_{K_{i,j}}\frac{\ms}{2}\overline{\left(v-\us\right)^2}\left(\d{\fs}{t}\right)_c\dxdv &= -\int_{x_{i-1/2}}^{x_{i+1/2}}\ms\nusr\sum_j\left[\left(\vcen_j-\us\right)\vtsr^2\hat{\fs}\Big|_{v_{j-1/2}}^{v_{j+1/2}} \right. \\
&\left.\quad+ \int_{v_{j-1/2}}^{v_{j+1/2}} \left(\vcen_j-\us\right)\left(v-\usr\right)\fs\dv\right]\dx,
\end{eqnal}
having used the continuity of $\overline{\left(v-\us\right)^2}\Gs$, its boundary conditions, equation~\ref{eq:vSqBar_x} and $\partial^2\vSqBar/\partial v^2=0$. Carry out the velocity-space integrals on the right as well as the sum over $j$ in order to land at
\begin{eqnal}
&\sum_j\int_{K_{i,j}}\frac{\ms}{2}\overline{\left(v-\us\right)^2}\left(\d{\fs}{t}\right)_c\dxdv \\
&\quad= \int_{x_{i-1/2}}^{x_{i+1/2}}\ms\nusr\left[\vtsr^2\left(\MZstars - \vcen_j\fs(v_{j\pm1/2})\Big|_{\jmin}^{\jmax} + \us\hat{\fs}\Big|_{v_{\jmin-1/2}}^{v_{\jmax+1/2}}\right) \right. \\
&\left.\qquad+ \usr\left(\MOstars-\us\MZs\right) - \left(\MTstars - \us\MOs\right)\right]\dx.
\end{eqnal}
Therefore when using $p=1$ bases we enforce the equality of the relaxation rates with
\begin{eqnal}
&\ms\nusr\left[\usr\left(\MOstars-\us\MZs\right) + \vtsr^2\left(\MZstars - \vcen_j\fs(v_{j\pm1/2})\Big|_{\jmin}^{\jmax} + \us\hat{\fs}\Big|_{v_{\jmin-1/2}}^{v_{\jmax+1/2}}\right) \right. \\
&\quad-\mr\nurs\left[\urs\left(\MOstarr-\ur\MZr\right) + \vtrs^2\left(\MZstarr - \vcen_j\fr(v_{j\pm1/2})\Big|_{\jmin}^{\jmax} + \ur\hat{\fr}\Big|_{v_{\jmin-1/2}}^{v_{\jmax+1/2}}\right) \right. \\
&\quad\doteq \ms\nusr\left(\MTstars - \us\MOs\right) - \mr\nurs\left(\MTstarr - \ur\MOr\right) \\
&\quad+ \frac{\alphae}{\MZs\MZr}\left[\mr\MZs\left(\MTr-\uri\MOir\right)-\ms\MZr\left(\MTs-\usi\MOis\right) \right.\\
&\left.\quad+\frac{\mr-\ms}{2}\left(\usi-\uri\right)\left(\MZr\MOis-\MZs\MOir\right)\right].
\end{eqnal}

\subsection{Gyroaveraged p=1 energy conservation} \label{sec:pOenerConservGK}

Energy conservation with $p=1$ basis functions is also possible with the gyroaveraged operator in equation~\ref{eq:gkLBO}. The discretization and calculation of the cross-primitive moments follows in the vein of that explained in section~\ref{sec:discreteGkLBO}, although this time we must consider the projection of the $\vpar^2$ onto the $p=1$ as was done for the non-gyroaveraged operator in the previous section. Substituting $\psi_\ell=\ms\vparSqBar/2$ into the weak form of the gyroaveraged collision operator, summing over velocity-space cells and species we obtain the following constraint
\begin{eqnal}
&\ms\nusr\left\{\uparsr\MOparstars + \vtsr^2\left[\MZstars+2\MZs-\frac{2\pi}{\ms}\left(\int\vparcenj\fs(v_{\parallel j\pm1/2})\Big|_{\jmin}^{\jmax}\dmu \right.\right.\right. \\
&\left.\left.\left.\quad+2\int\mu\fs\Big|_{\mumin}^{\mumax}\dvpar\right)\right]\right\} \\
\quad+&\mr\nurs\left\{\urs\MOparstarr + \vtrs^2\left[\MZstarr+2\MZr-\frac{2\pi}{\mr}\left(\int\vparcenj\fr(v_{\parallel j\pm1/2})\Big|_{\jmin}^{\jmax}\dmu \right.\right.\right. \\
&\left.\left.\left.\quad+2\int\mu\fr\Big|_{\mumin}^{\mumax}\dvpar\right)\right]\right\} \doteq \ms\nusr\MTstars + \mr\nurs\MTstarr.
\end{eqnal}

In addition to energy conservation the gyroaveraged LBO-G requires the discrete thermal relaxation rate of the operator, which we must re-calculate assuming $p=1$ basis functions. Multiplying the discrete gyroaveraged LBO by $\psi_\ell=\ms\overline{\left(v-\upars\right)^2}/2+\mu B$, integrating over phase-space and summing over velocity space cells we obtain the following discrete relaxation rate:
\begin{eqnal} \label{eq:gkLBOthermalRateDiscrete}
&\sum_{j,k}\int_{K_{i,j,k}}\left[\frac{\ms}{2}\overline{\left(\vpar-\upars\right)^2}+\mu B\right]\left(\pd{\fs}{t}\right)_c\dx\,\dvpar\,\dmu = \\
&\quad\nusr\sum_{j,k}\left(\int_{\mu_{k-1/2}}^{\mu_{k+1/2}}\left\{\left[\frac{\ms}{2}\overline{\left(\vpar-\upars\right)^2}+\mu B\right]G_{\vpar\sind}-\ms\left(\vparcenj-\upars\right)\vtsr^2\hat{\fs}\right\}\Bigg|_{v_{\parallel j-1/2}}^{v_{\parallel j+1/2}}\dmu \right. \\
&\left.\quad+\int_{v_{\parallel j-1/2}}^{v_{\parallel j+1/2}}\left\{\left[\frac{\ms}{2}\overline{\left(\vpar-\upars\right)^2}+\mu B\right]G_{\mu\sind}-2\mu\ms\vtsr^2\hat{\fs}\right\}\Bigg|_{\mu_{k-1/2}}^{\mu_{k+1/2}}\dvpar \right. \\
&\left.\quad-\int_{K_{i,j,k}}\left[\ms\left(\vparcenj-\upars\right)\left(\vpar-\uparsr\right)\fs+B2\mu\fs-2\ms\vtsr^2\fs\right]\dx\,\dvpar\,\dmu\right), \\
\end{eqnal}
where $j(k)$ labels the cell along $\vpar$($\mu$), and we used the fact that $\ms\overline{\left(v-\upars\right)^2}/2+\mu B$ is linear in $\vpar$ and that its $\vpar$ derivative is $\ms\left(\vparcenj-\upars\right)$. In equation~\ref{eq:gkLBOthermalRateDiscrete} the $G_{\vpar\sind}$ and $G_{\mu\sind}$ are numerical fluxes~\citep{Francisquez2020}. Doing the velocity-space integrals, carrying out the sums over velocity space cells, using the continuity of $G_{\vpar\sind}$, $G_{\mu\sind}$ and $\hat{\fs}$ and the zero-flux BCs one obtains
\begin{eqnal}
&\sum_{j,k}\int_{K_{i,j,k}}\left[\frac{\ms}{2}\overline{\left(\vpar-\upars\right)^2}+\mu B\right]\left(\pd{\fs}{t}\right)_c\dx\,\dvpar\,\dmu = \\
&\quad\ms\nusr\int_{x_{i-1/2}}^{x_{i+1/2}}\left\{\vtsr^2\left[2\MZs+\MZstars-\frac{2\pi}{\ms}\left(\sum_k\int_{\mu_{k-1/2}}^{\mu_{k+1/2}}\vparcenj\fs(v_{\parallel j\pm1/2})\Big|_{\jmin}^{\jmax}\dmu \right.\right.\right.\\
&\left.\left.\left.\quad+2\sum_j\int_{v_{\parallel j-1/2}}^{v_{\parallel j+1/2}}\mu\fs\Big|_{\mumin}^{\mumax}\dvpar\right)\right]+\uparsr\left(\MOparstars-\upars\MZs\right)-\left(\MTstars-\upars\MOpars\right)\right\}\dx.
\end{eqnal}
Using this equation we can enforce the equality between the discrete thermal relaxation rates via
\begin{eqnal}
&\ms\nusr\left\{\uparsr\left(\MOparstars-\upars\MZs\right) + \vtsr^2\left[2\MZs+\MZstars \right.\right. \\
&\left.\left.\quad-\frac{2\pi}{\ms}\left(\sum_k\int_{\mu_{k-1/2}}^{\mu_{k+1/2}}\vparcenj\fs(v_{\parallel j\pm1/2})\Big|_{\jmin}^{\jmax}\dmu+2\sum_j\int_{v_{\parallel j-1/2}}^{v_{\parallel j+1/2}}\mu\fs\Big|_{\mumin}^{\mumax}\dvpar\right)\right]\right\} \\
&\quad-\mr\nurs\left\{\uparrs\left(\MOparstarr-\uparr\MZr\right) + \vtrs^2\left[2\MZr+\MZstarr\right.\right. \\
&\left.\left.\quad-\frac{2\pi}{\mr}\left(\sum_k\int_{\mu_{k-1/2}}^{\mu_{k+1/2}}\vparcenj\fr(v_{\parallel j\pm1/2})\Big|_{\jmin}^{\jmax}\dmu+2\sum_j\int_{v_{\parallel j-1/2}}^{v_{\parallel j+1/2}}\mu\fr\Big|_{\mumin}^{\mumax}\dvpar\right)\right]\right\} \\
&\quad\doteq \ms\nusr\left(\MTstars - \upars\MOpars\right) - \mr\nurs\left(\MTstarr - \uparr\MOparr\right) \\
&\quad+ \frac{\alphae}{\MZs\MZr}\left[\mr\MZs\left(\MTr-\uparr\MOparr\right)-\ms\MZr\left(\MTs-\upars\MOpars\right) \right.\\
&\left.\quad+\frac{\mr-\ms}{2}\left(\upars-\uparr\right)\left(\MZr\MOpars-\MZs\MOparr\right)\right].
\end{eqnal}

\bibliographystyle{jpp}

\bibliography{main}

\end{document}